\documentclass[]{pasj01}
\draft

\Received{}
\Accepted{}
 
\usepackage{graphicx}
 \usepackage{txfonts}
\usepackage{subfigure}

\begin{document}

\title{\textbf{HCN 3-2 survey towards a sample of local galaxies}}
\author{Fei \textsc{Li}\altaffilmark{1,2*},
Junzhi \textsc{Wang}\altaffilmark{1,3*},
Min \textsc{Fang}\altaffilmark{4},
Qing-hua \textsc{Tan}\altaffilmark{5},
Zhi-Yu \textsc{Zhang}\altaffilmark{6},
Yu \textsc{Gao}\altaffilmark{5,7},
Shanghuo \textsc{Li}\altaffilmark{1,2,8}}%
\altaffiltext{1}{Shanghai Astronomical Observatory, Chinese Academy of Sciences,  80 Nandan Road, Shanghai, China, 200030}
\altaffiltext{2}{University of the Chinese Academy of Science, No.19A Yuquan Road, Beijing, China, 100049}
\altaffiltext{3}{Key Laboratory of Radio Astronomy, Chinese Academy of Sciences,  Nanjing, 210008,  China}
\altaffiltext{4}{Department of Astronomy, University of Arizona, 933 North Cherry Avenue, Tucson, AZ 85721, USA}
\altaffiltext{5}{Purple Mountain Observatory, Chinese Academy of Sciences, 8 Yuanhua Road, Nanjing 210034, China}
\altaffiltext{6}{School of Astronomy and Space Science, Nanjing University, 163 Xianlin Avenue, Nanjing 210023, China}
\altaffiltext{7}{Department of Astronomy, Xiamen University, Xiamen, Fujian 361005, China}
\altaffiltext{8}{Harvard-Smithsonian Center for Astrophysics, 60 Garden Street, Cambridge, MA 02138, USA}
\email{lifei@shao.ac.cn}
\email{jzwang@shao.ac.cn}


\maketitle

\begin{abstract}
We present observations of HCN 3-2 emissions towards 37 local galaxies using 10-m Submillimeter Telescope (SMT).  HCN 3-2 emission is detected in 23 galaxies. The correlation of infrared luminosity ($L_{\rm IR}$) and the luminosity of HCN 3-2 line emission measured in our sample is fitted with a slope of 1.11 and correlation coefficient of 0.91, which follows the linear correlation  found in other dense gas tracers in the literatures. Although molecular gas above a certain volume density threshold (i.e., $n_{\rm H_2}\geq$ 10$^4$ cm$^{-3}$) statistically gave similar relation with infrared luminosity, the large scatter of HCN 3-2/HCN1-0 ratios for galaxies with different $L_{\rm IR}$ indicates that dense gas masses estimated from the line luminosities of only one transition of dense gas tracers should be treated with caution for individual galaxies.
\end{abstract}

\section{Introduction} \label{sec:intro}

Observational evidence of dense gas tracers has suggested that star formation is closely related to the dense cores of molecular clouds, in the Milky Way and in the extragalactic domain \citep{Kennicutt2012,Lada2012}. The star formation law, the relation between star formation rate (SFR) and gas mass, is a fundamental tool to study star formation processes and to understand galaxy formation and evolution. The Kennicutt-Schmidt (K-S) law \citep{Kennicutt1998} formulate the global surface densities of SFR traced by H$_{\alpha}$ is closely tied up with the total gas surface densities, traced by CO and HI 21 cm line, with an index of N = 1.4$\pm$0.15.

CO $J =1 \rightarrow 0$ and  $2 \rightarrow 1$ (hereafter 1-0 and 2-1) can trace the bulk of molecular  gas content of low to medium density, while transitions from high dipole-moment molecules only trace dense molecular gas. 
A tight linear relation has been found between the infrared luminosity ($L_{\rm IR}$) which traces the SFR, and HCN luminosity ($L_{\rm HCN 1-0}$) which  traces mass of dense molecular gas,  in galaxies by \citep{Gao2004b}. This relation was extended to the Milky Way dense cores \citep{Wu2005,Shimajiri2017} and high-$z$ galaxies  \citep{Gao2007}.

However, the linear correlation of $L_{\rm IR}$-$L'_{\rm dense}$ was still debated. The theoretical studies of the star formation predict that decreasing slopes against increasing critical densities ($n_{\rm crit}$) \citep{Krumholz2007,Narayanan2008}. Some observations of HCN 3-2  show that the slope of $L_{\rm  IR}$-$L'_{\rm  HCN(3-2)}$ is significantly below unity \citep{Bussmann2008,Juneau2009}, while a slightly super-linear IR-HCN 1-0 luminosity relation is found in \cite{Gracia-Carpio2008}, \cite{Garca-Burillo2012} and\cite{Saito2018}toward the local luminous and ultra-luminous infrared galaxies (LIRGs and ULIRGs). Note that the infrared luminosity in \cite{Bussmann2008} is the total infrared luminosity of an entire galaxy, while HCN 3-2 emissions are only from central part of the nearby  galaxies. On the other hand,  the sample of \cite{Gracia-Carpio2008} and \cite{Garca-Burillo2012} are just for LIRGs, ULIRGs and high-$z$ galaxies, with very limited infrared luminosity range.

Dense gas tracers with higher critical densities: CS 5-4 in \cite{Wang2011}; HCN 4-3, CS 7-6 in \cite{Zhang2014} and HCN 4-3 and HCO$^+$ 4-3 in \cite{Tan2018},  were all found to follow the linear slope. Liner slope was also found for $J$ = 6-5 transition of CO \citep{Liu2015,Kamenetzky2016}. Thus, re-checking the relation of $L_{\rm  HCN 3-2}$ and $L_{\rm IR}$ are necessary for understanding dense tracers and star formation in galaxies.  

In this paper, we describe the observations and data reduction in Section 2, and the main results are presented in Section 3. In Section 4, we present the analysis and discussion. The final conclusions are summarized in Section 5.

\section{Observations and data reduction}
\label{sec:obs}

A sample of 37 local galaxies from the far-IR survey of Infrared Astronomical Satellite (IRAS)
\citep{Sanders2003} are selected. We carry out the new observations for HCN 3-2. There are 33 sources, 
with 60 $\mu$m flux densities greater than 20 Jy. And the other 4 sources with 60 $\mu$m flux less than 20 Jy are selected from \cite{Gao2004a} with the strong HCN 1-0 emission. 
 

\subsection{HCN 3-2 observations with the SMT 10-m telescope}

The observations of HCN 3-2 toward 37 local galaxies,  were carried out with the 10 m SMT telescope on Mt. Graham, AZ, between December 2015 and February 2016. Then, six sources (M 82, NGC 3504,  NGC 3079, NGC 4418, NGC 6240, and NGC 6946)  with strong HCN 3-2 emission were selected to observe H$^{13}$CN 3-2, which were shown in another paper (Li et al. 2020). The beam size of SMT is about 28$''$ at 265.886 GHz for HCN 3-2. Lower sideband (LSB) of the 1.3 mm ALMA Band 6 receiver with dual-polarization sideband-separating mixers was used for this observation. The backends employed were Forbes Filter Bank system with a 1 MHz frequency spacing and 1024 channels for each polarization. The channel width corresponds to a velocity separation of $\sim$ 1.2 km s$^{-1}$ at the observing frequencies. The beam-switching mode with a subreflecter throw of 2$^{'}$ was used for all observations. Telescope pointing and focus based on Jupiter and Mars were checked every two hours. Typical system temperatures were less than 240 K for all observations.  The antenna temperature $T_a^{\ast}$ was converted to main beam temperature $T_{mb}$ using $T_{mb}$=$T_a^{\ast}$/$\eta_b$, where $\eta_b$ is  the corrected beam efficiency. The integration time of HCN 3-2  was about 30 minutes for each source.

\subsection{Data reduction}

The  basic parameters of the samples are listed in Table \ref{tabel 1}.  The CLASS package, which is a part of the GILDAS\footnote{http://www.iram.fr/IRAMFR/GILDAS} software, was used for data reduction. Firstly, we checked each spectrum and 
qualified spectra by their baseline flatness, standing wave, system temperature, etc. Then, we averaged all reliable spectra into one spectrum for each source. First-order polynomial baseline was fitted and subtracted from the averaged spectrum for each source. 
The averaged spectra are smoothed to velocity resolutions of $\sim$ 20 - 40 km s$^{-1}$. The velocity-integrated intensities of these line are derived from the Gaussian fit to the spectra, or integrated over a defined window if the line profiles significantly deviate from a Gaussian profile. 

\begin{table}
\renewcommand\arraystretch{0.65}	
 \caption{List of targets}\label{tabel 1}
  \begin {center} 
  \renewcommand\tabcolsep{3.5pt}
		\begin{tabular}{l l l l l l l l l l l l l l l}
   	         \hline
	         \hline
	         {Source} & {RA} & {DEC}  & {F$_{60\mu m}$}  &{cz}   &{Distance} & {$D_{25}$} &{Beam size} &{log L$_{IR}$} & {I$_{HCN}$}&{log L$_{HCN}$} &  \\
                             & (J2000)  & (J2000)  &(Jy)              &(km\,s$^{-1}$) & (Mpc) &(kpc) &  (kpc) &(L$\odot$) &(K km\,s$^{-1}$) &(K km\,s$^{-1}$)  \\
	         \hline
	          \hline

NGC2146$^\mathrm{a}$   &  06:18:37.7      &    78:21:25    &       146.69 &     893     &   16.47    &  28.7& 2.2  &    10.5 $\pm$   0.1  & 0.51   $\pm$  0.12 & 6.5 $\pm$0.1\\   
NGC2798   &  09:17:22.8    &     41:59:59   &        20.60 &     1726  &   27.84  &  21.1 & 3.8 &    10.8 $\pm$    0.1   & $<$ 0.55   & $<$ 6.93 \\    
NGC2903$^\mathrm{a}$	&  09:32:10.5   &21:30:05      &    60.54  & 566  &  8.26  & 30.3 &  1.1 &    9.2$\pm$  0.2     &   0.66$\pm$ 0.10  &  6.0 $\pm$ 0.1\\
NGC3031	&  09:55:33.6   & 69:03:56     &    44.73 &  -34   &  3.63   & 28.4 & 0.5 &   8.2$\pm$   0.02   &    $<$ 0.30   &  $<$ 4.90  \\
NGC3034$^\mathrm{a}$ &  09:55:53.1	&69:40:41      &    1480.42   & 203  &   3.63  &  11.8  & 0.5   &    10.0$\pm$ 0.03      &      2.11  $\pm$  0.15  &6.3 $\pm$ 0.02 \\
NGC3079$^\mathrm{a}$ &  10:01:57.9   &55:40:51      &    50.67 & 1116     &   18.19 & 41.8 & 2.5 &      10.7$\pm$ 0.1  &    1.18   $\pm$ 0.27& 6.9 $\pm$ 0.1  \\
NGC3310 &  10:38:46.2   &53:30:08      &    34.56  & 1060  &  19.81 &  17.9 & 2.7   &    10.1$\pm$ 0.1   &   $<$0.39   &  $<$ 6.49\\
NGC3351   & 10:43:58.1 &     11:42:10  &     19.66    &     778   &   9.99  &8.9 &  1.4 &      9.8$\pm$  0.1 &   0.97  $\pm$ 0.21&  6.3 $\pm$  0.1   \\   
NGC3504 &   11:03:11.1    & 27:58:22        &  21.43    & 1525    &    27.07  & 21.3 & 3.7  &   10.5 $\pm$ 0.1   &  0.94  $\pm$  0.16   &  7.2 $\pm$  0.1 \\
NGC3521 &   11:05:49.2    &-00:02:15        &  49.19    & 801   &   6.84 & 21.9 & 0.9  &  9.1$\pm$  0.1   &  0.53  $\pm$  0.15   & 5.7 $\pm$  0.1 \\
NGC3627   &     11:20:14.9   &     12:59:30   &   66.31      &    727  &  10.04 &  26.6  & 1.4  &  9.8 $\pm$ 0.1  &   0.70  $\pm$   0.20  & 6.2 $\pm$  0.1 \\  
NGC3628$^\mathrm{a}$   &     11:20:17.0   &     13:35:23   &    54.8      &      843    & 10.04  & 43.2 & 1.4   &     10.2  $\pm$ 0.1  &  0.32   $\pm$  0.06 & 5.8 $\pm$  0.1 \\   
NGC4088 &	12:05:35.1	  & 50:32:24	       &  26.77    & 757   &     13.37 & 22.6 & 1.8 &      9.0 $\pm$  0.4  &   0.32 $\pm$   0.10   &  6.1 $\pm$ 0.1 \\
NGC4102 &  12:06:23.6   &52:42:36      &    46.85  & 846   &   16.89  & 13.3 & 2.3 &    10.3 $\pm$  0.1 &  0.78 $\pm$   0.22   &  6.3  $\pm$  0.1   \\
NGC4194 &    12:14:08.7    &54:31:40        &  23.20    & 2051   &  40.33 & 21.1 & 5.5 &     11.1  $\pm$  0.1 &   $<$ 0.68  & $<$ 7.35 \\
NGC4254	&  12:18:51.0   &14:24:50      &    7.46  & 2407   &  15.29  & 24.0  &2.1 &    9.8  $\pm$   0.1   &    0.74 $\pm$   0.18   &  6.5 $\pm$ 0.1 \\
NGC4303	&  12:21:55.4   &04:28:24      &    37.27  & 1566   &  15.29 & 28.9 &  2.1 &      9.6 $\pm$   0.03  &     $<$0.29   & $<$ 6.15  \\
NGC4321 &  12:22:53.9   &15:49:22      &     26.00  &  1571   &   15.20 &32.7   &2.1  &     10.1 $\pm$  0.1  &    $<$0.46 &  $<$6.34  \\
NGC4414$^\mathrm{a}$	&  12:26:26.9   &31:13:24      &    29.55  & 716    &  17.68 & 18.5 & 2.4 &    9.3 $\pm$  0.1   & $<$  0.65  & $<$ 6.62  \\ 
NGC4418 &  12:26:54.7   &-00:52:42     &    43.89   & 2179     &   31.90  & 14.8  &  4.3  &      11.0 $\pm$   0.1  &   0.86 $\pm$   0.21   &  7.3   $\pm$  0.1 \\
NGC4490 &  12:30:34.9   &41:38:47      &    46.92   &  565   &  10.48  & 19.2 &  1.4&   8.7 $\pm$  0.5      &  0.30  $\pm$   0.06  &  5.8  $\pm$ 0.1 \\
NGC4501   &      12:31:57.6  &    14:25:20   &       19.68  &     2281 & 15.29 & 30.7 & 2.1  &  9.6$\pm$ 0.1  &  $<$1.5   & $<$ 6.85 \\   
NGC4527 &  12:34:09.9   &02:39:04      &    31.40  &  1736  &  15.29  & 27.6  & 2.1 &    10.1 $\pm$   0.11 &  $<$ 0.44   &  $<$ 6.32  \\
NGC4536 &  12:34:28.5   &02:11:08      &    30.26  & 1808   &14.92  & 33.0  &  2.0  &     10.3 $\pm$  0.1  &   0.92  $\pm$  0.13  &  6.6  $\pm$  0.1  \\
NGC4568/7 &     12:36:33.7   &   11:14:32    &   20.81      &      2255   &  15.3 & 18.9  & 2.1 &     10.16 $\pm$  0.1 & $<$  0.52  &  $<$ 6.40 \\   
NGC4631   &     12:42:07.1   &     32:32:33  &   85.40      &     606    &  7.73  &  34.9 &1.0  &    9.7  $\pm$  0.6  &  $<$ 0.37 &  $<$ 5.70 \\   
NGC4736   &       12:50:52.9 &    41:07:15   &   71.54      &      308   &   4.83   & 15.7  & 0.7  &    9.3     $\pm$    0.1 &    0.78  $\pm$ 0.22 &  5.6 $\pm$  0.1   \\   
NGC4826 &   12:56:42.6     & 21:41:05       &   36.70   &  408   &  3.09    &  9.0  & 0.4 &    9.1   $\pm$  0.1 &  $<$  0.50 &  $<$ 5.00 \\
NGC5055   &      13:15:49.5  &    42:01:39   &      40      &     484  &  7.96   & 29.2  & 1.1  &     9.6 $\pm$   0.1  &  $<$  0.58  & $<$ 5.90   \\   
NGC5194 &  13:29:53.5   &47:11:42      &    97.42  & 463  &   8.63    & 28.1 &1.2  &   9.5  $\pm$ 0.1      &   0.7  $\pm$  0.23   &  6.0$\pm$ 0.1  \\
NGC5253 &  13:39:55.2   &-31:38:21     &    29.84    &  407   & 3.15 & 4.6  &  0.4   &    8.8  $\pm$  0.1  &   $<$ 0.41 & $<$ 4.90 \\
NGC5457 &  14:03:09.0   &54:21:24      &    88.04   &  241   &   6.70   & 56.1  & 0.9 &   8.7 $\pm$ 0.1  &  0.45   $\pm$  0.18 & 5.6 $\pm$ 0.2\\
NGC5713 &   14:40:10.9     & -00:17:22      &   22.10   &  1899   &  26.74  & 21.8  & 3.6 &     10.6    $\pm$   0.1 &  0.75  $\pm$  0.24 & 7.0 $\pm$ 0.1  \\
NGC5775 &   14:53:58.0     & 03:32:32       &   23.59   &  1681   &   26.34   & 32.2 & 2.6 &   9.6    $\pm$     0.5  &   0.33  $\pm$  0.09    &  6.7  $\pm$ 0.1  \\
CGCG049-057 & 15:13:12.7   & 07:13:30       &   21.89   &   3897   &   59.06  & 6.9 &  8.0   &     11.5 $\pm$      0.1 &   0.96   $\pm$ 0.21   &  7.8 $\pm$ 0.1\\
NGC6240$^\mathrm{a}$   &    16:52:58.9    &    02:24:03    &   22.94      &      7200  &  103.86 & 63.4  & 14.1  &     11.8   $\pm$  0.1  &      1.7   $\pm$     0.29   & 8.6 $\pm$  0.1 \\    
NGC6946   &      20:34:52.6  &   60:09:12 &   129.78     &      40  &    5.32  & 17.8  & 0.7   &     9.3$\pm$  0.02   &     1.35   $\pm$   0.31  & 5.9 $\pm$ 0.1 \\   
 \hline
 Mrk231$^\mathrm{b}$     &     12:56:14.2   &   56:52:25     &  30.8        &    12139 &  171.84 & -&- &   12.4  $\pm$  0.1 & 38.2 $\pm$0.4$^\mathrm{e}$ &  8.6 $\pm$ 0.01 \\   
 Mrk273$^\mathrm{c}$     &   13:44:42.1  &     55:53:13   &       22.51  &     11326  &  154.71 & -&- & 12.1 $\pm$ 0.1   &  29.7 $\pm$ 0.6$^\mathrm{e}$ &  8.6 $\pm$  0.01 \\   
 SDP.9$^\mathrm{d}$       &   09:07:40.0   &  -00:41:59.8  &  $\ -$       &  $\ -$    &  6654.93  & -&- &  13.8 $\pm$  0.01  &   0.66  $\pm$  0.11$^\mathrm{e}$  &  9.7$\pm$ 0.1 \\
 SDP.11$^\mathrm{d}$     &   09:10:43.1   &   -00:03:22.8  & $\ - $       &  $\ -$   &   7546.48  & -&- &  13.8 $\pm$  0.01  &  0.54 $\pm$ 0.08$^\mathrm{e}$   & 9.7 $\pm$  0.1\\
 
	         \hline	  
		\end{tabular}
  \end{center}
 Notes.  The fluxes are from the Gaussian fitting.  $^\mathrm{a}$ these seven sources overlap with the sample in \cite{Bussmann2008}. Mrk 231, Mrk 273,  SDP.9 and  SDP.11 are from the literatures: $^\mathrm{b}$\citep{Aalto2015}, 
 $^\mathrm{c}$\citep{Aladro2018}, $^\mathrm{d}$\citep{Oteo2017}. $^\mathrm{e}$ the flux of HCN is presented with unit of Jy km s$^{-1}$.
 The 60 $\mu$m flux densities and distance of these galaxies are from \cite{Sanders2003}. Optical diameter (D$_{25}$) is from NED \footnotemark[2].

 \end{table}
 \footnotetext[2]{The NASA/IPAC Extragalactic Database (NED)}


\begin{figure*}[ht]


  \subfigure 


\includegraphics[height=1.3in,width=2.2in]{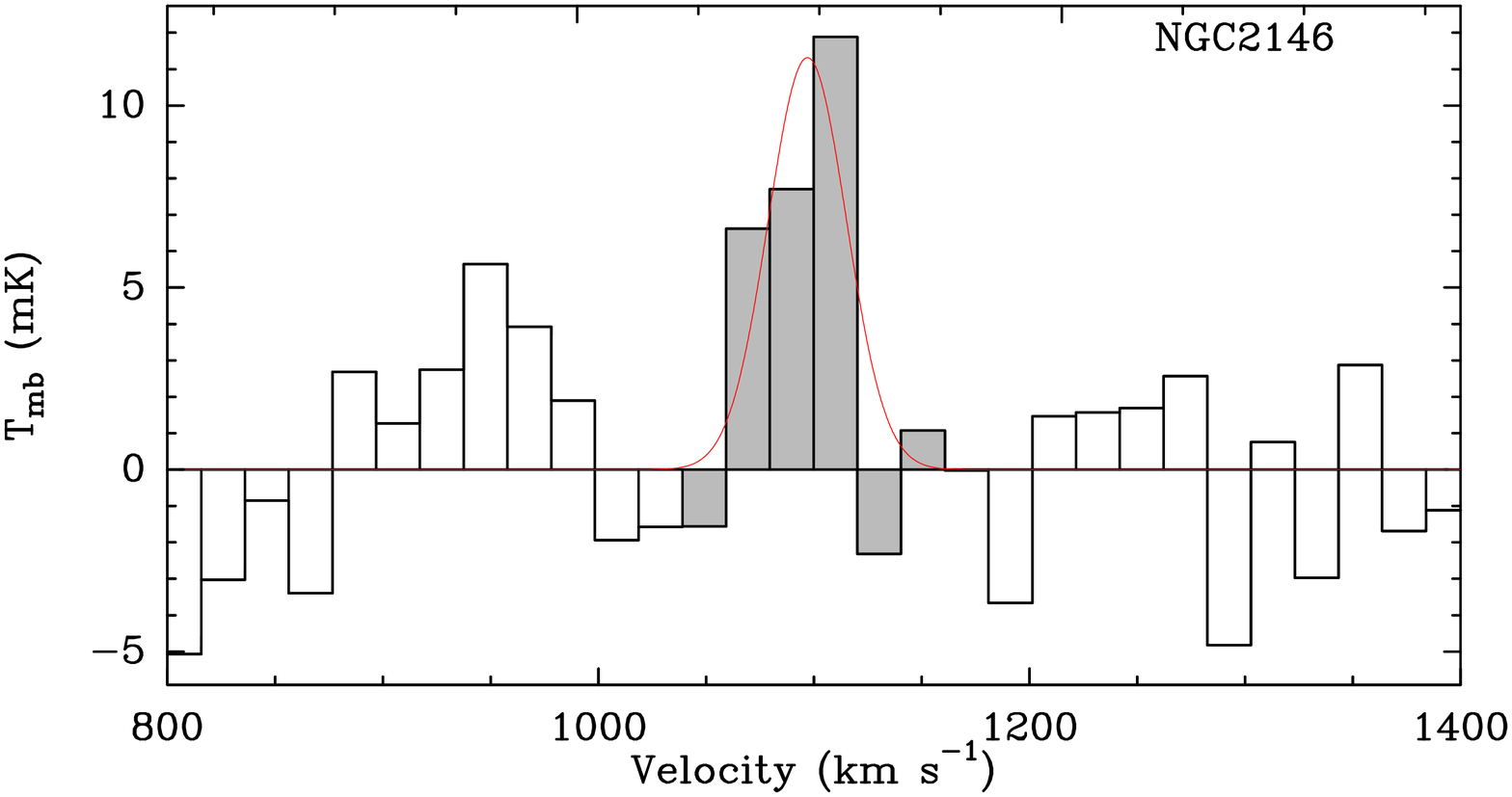}
\includegraphics[height=1.3in,width=2.2in]{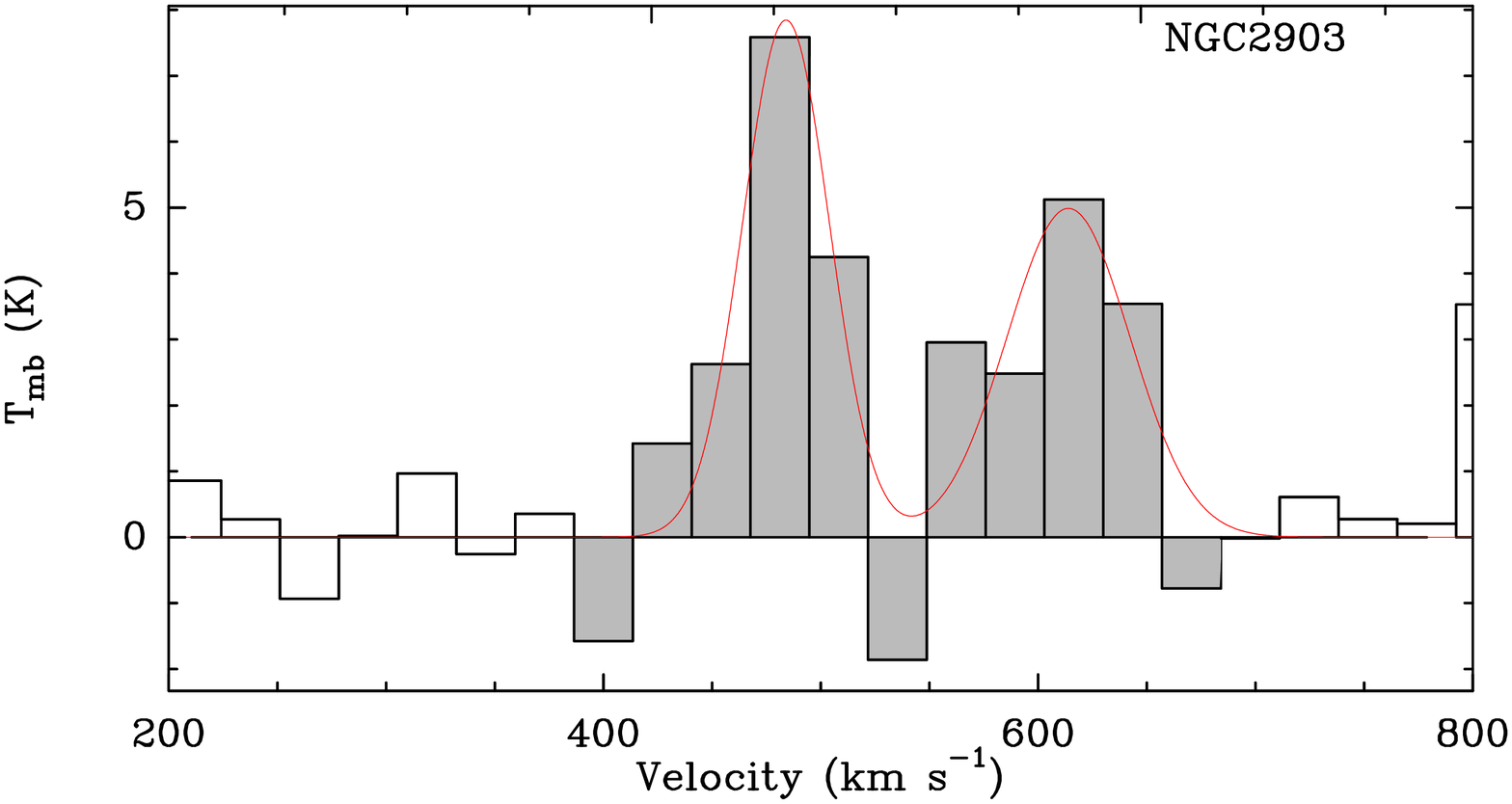}
\includegraphics[height=1.3in,width=2.2in]{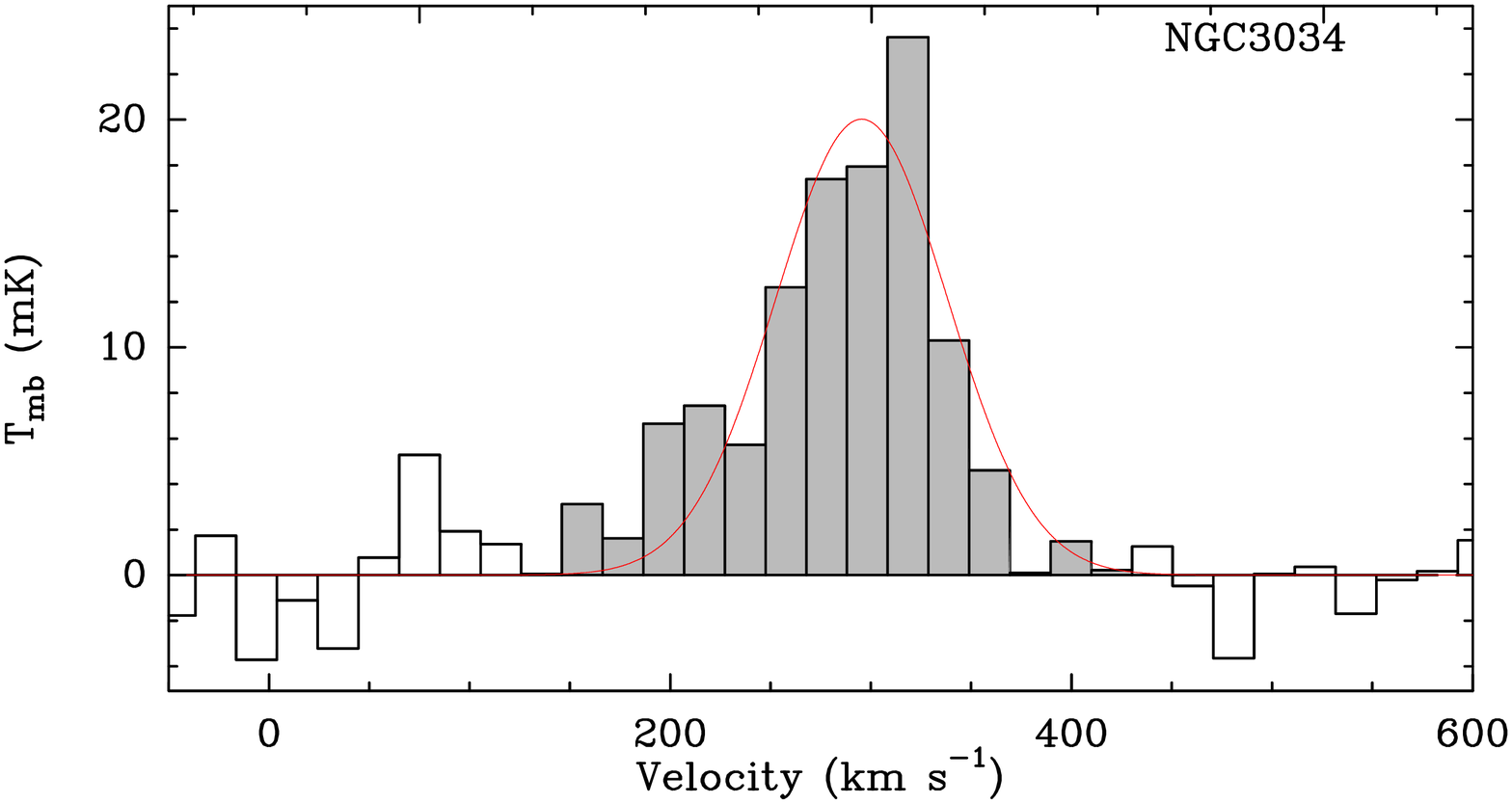}
\includegraphics[height=1.3in,width=2.2in]{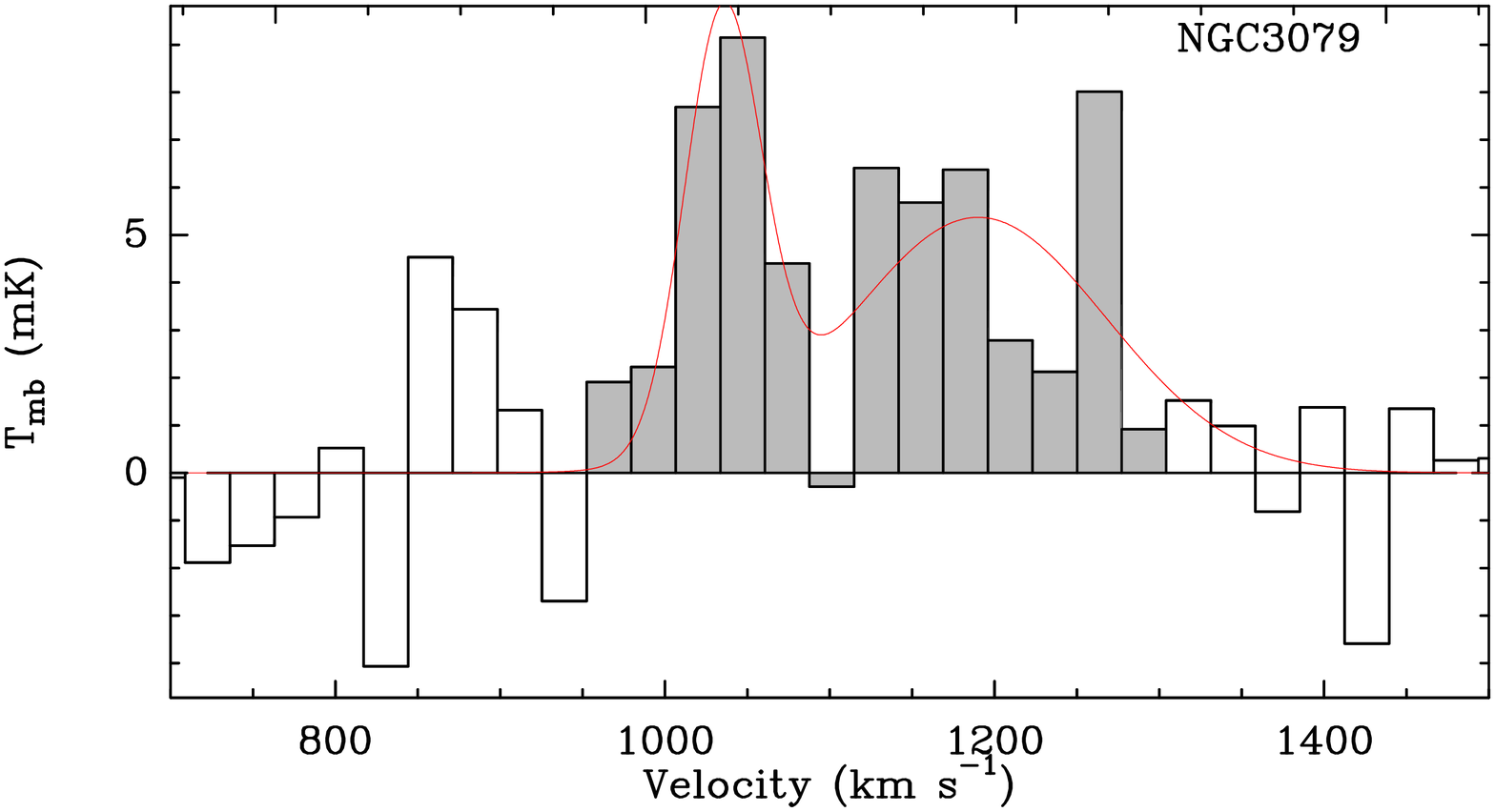}
\includegraphics[height=1.3in,width=2.2in]{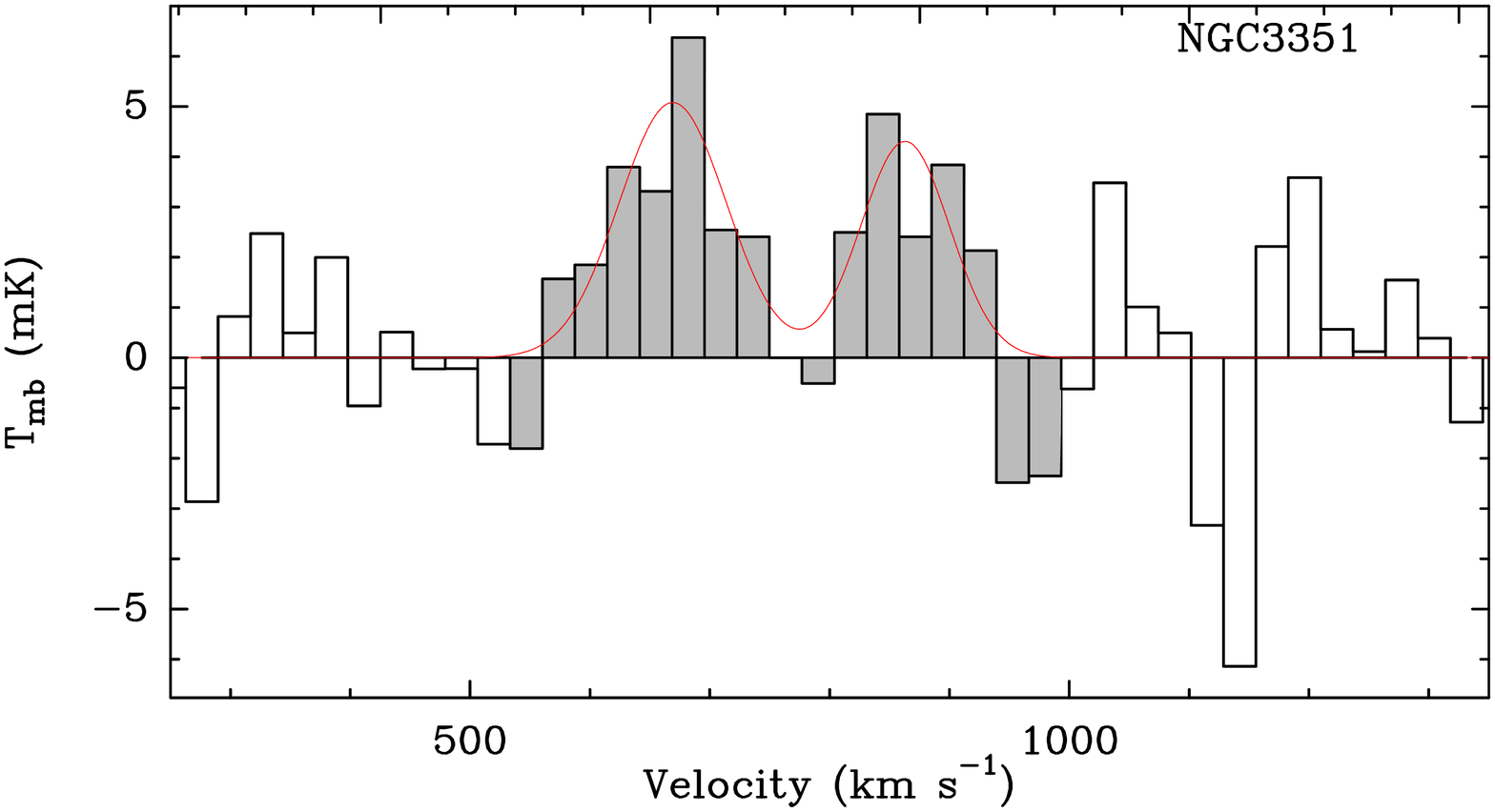}
\includegraphics[height=1.3in,width=2.2in]{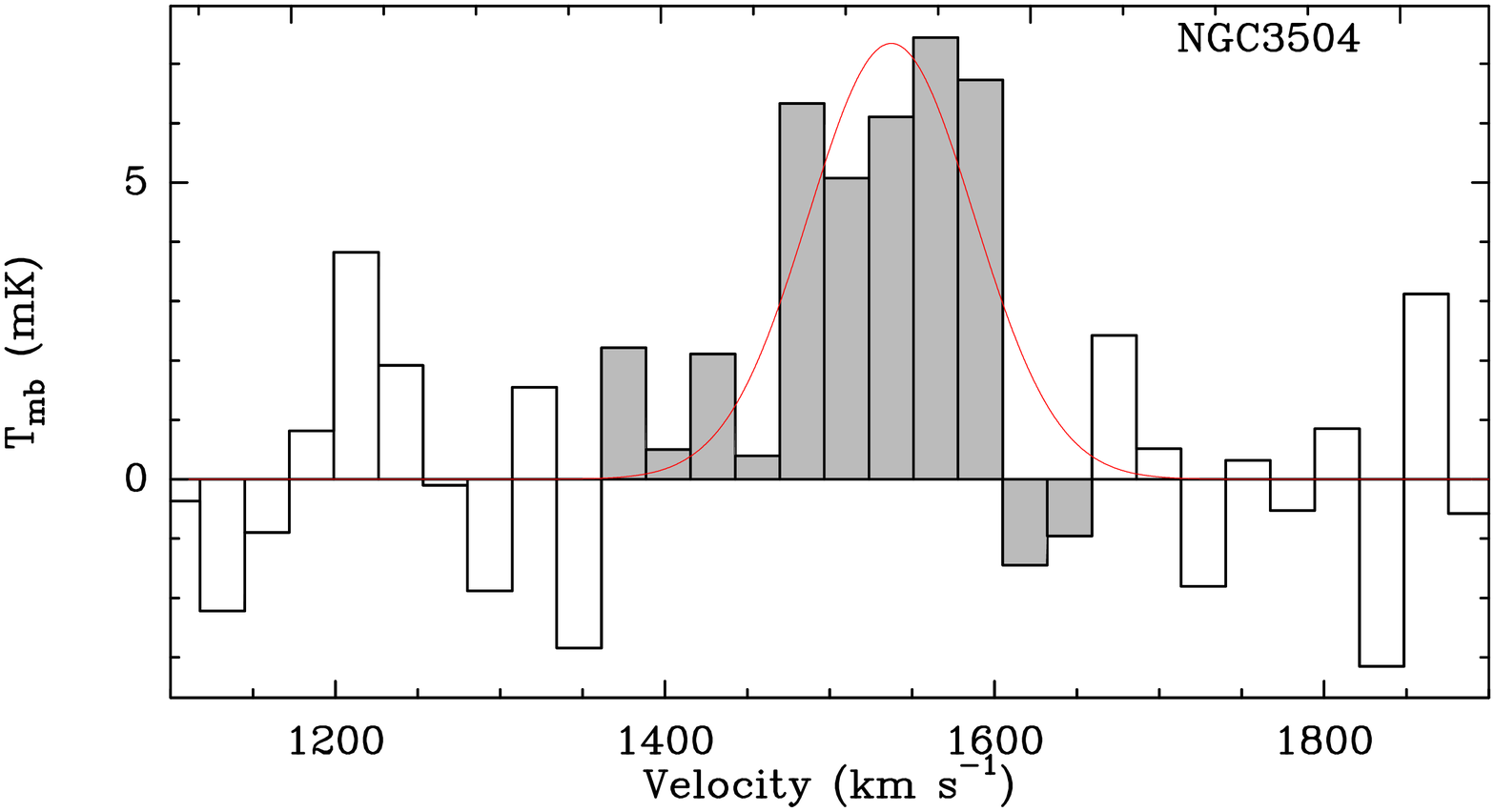}
\includegraphics[height=1.3in,width=2.2in]{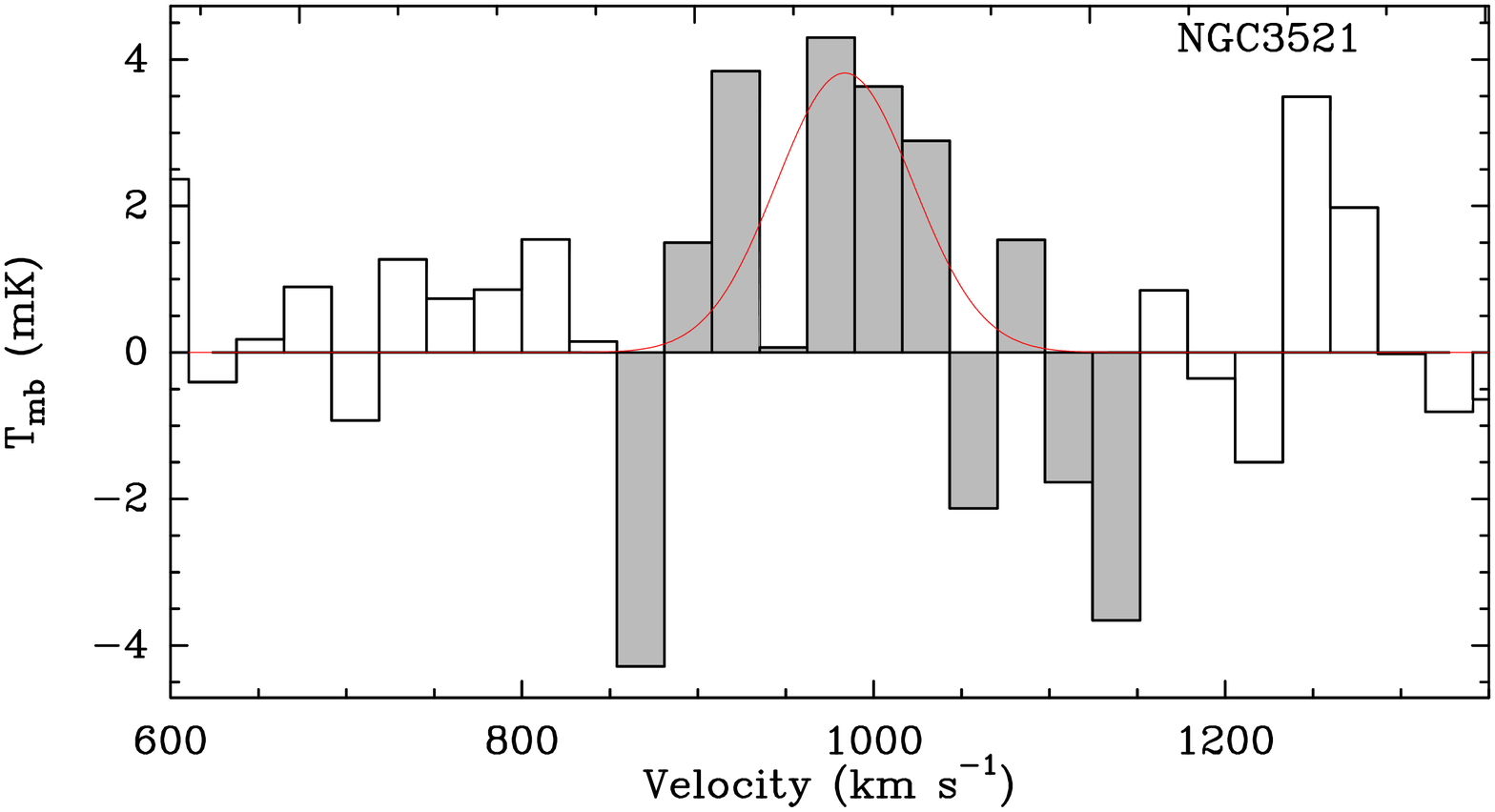}
\includegraphics[height=1.3in,width=2.2in]{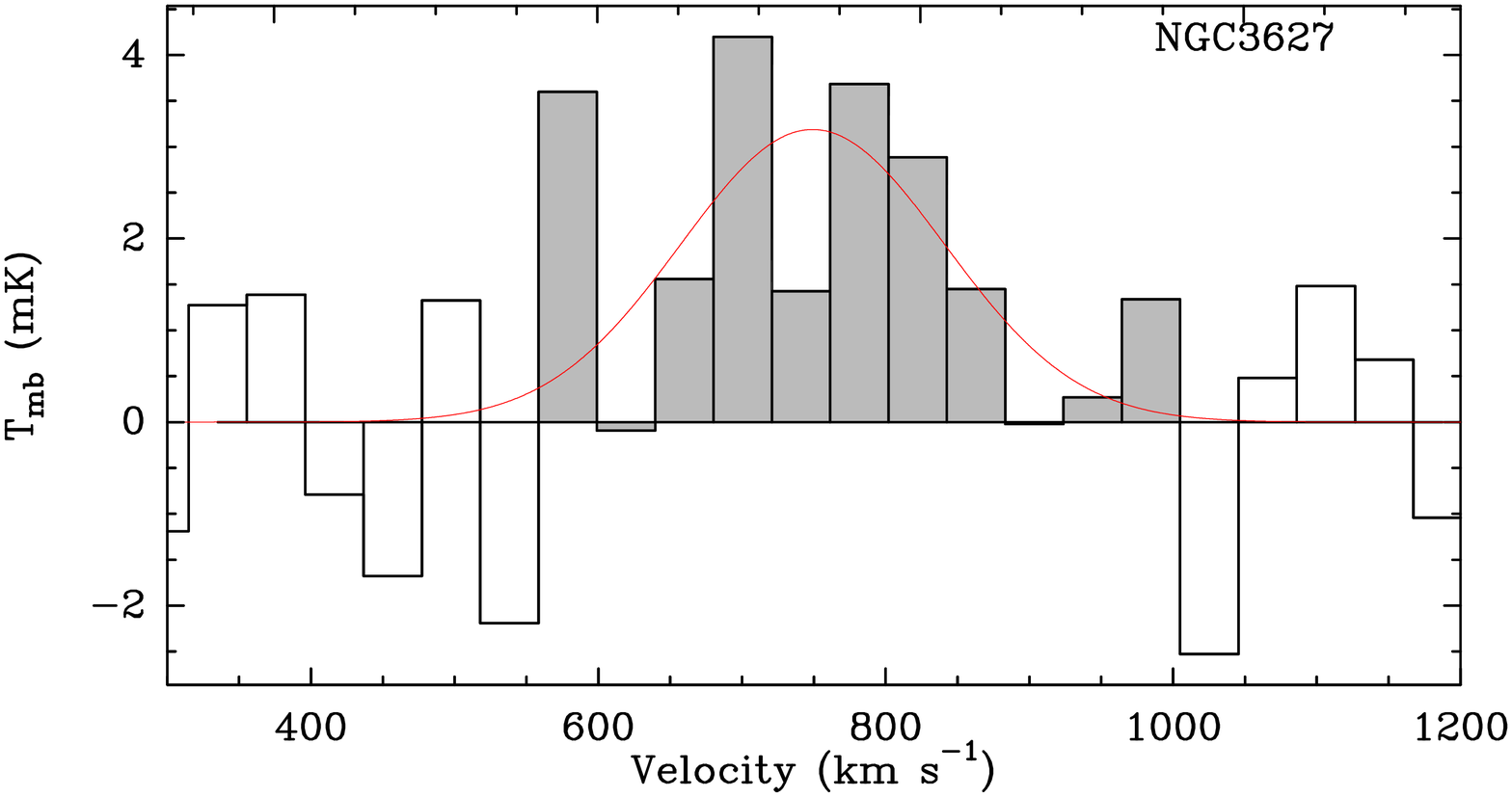}
\includegraphics[height=1.3in,width=2.2in]{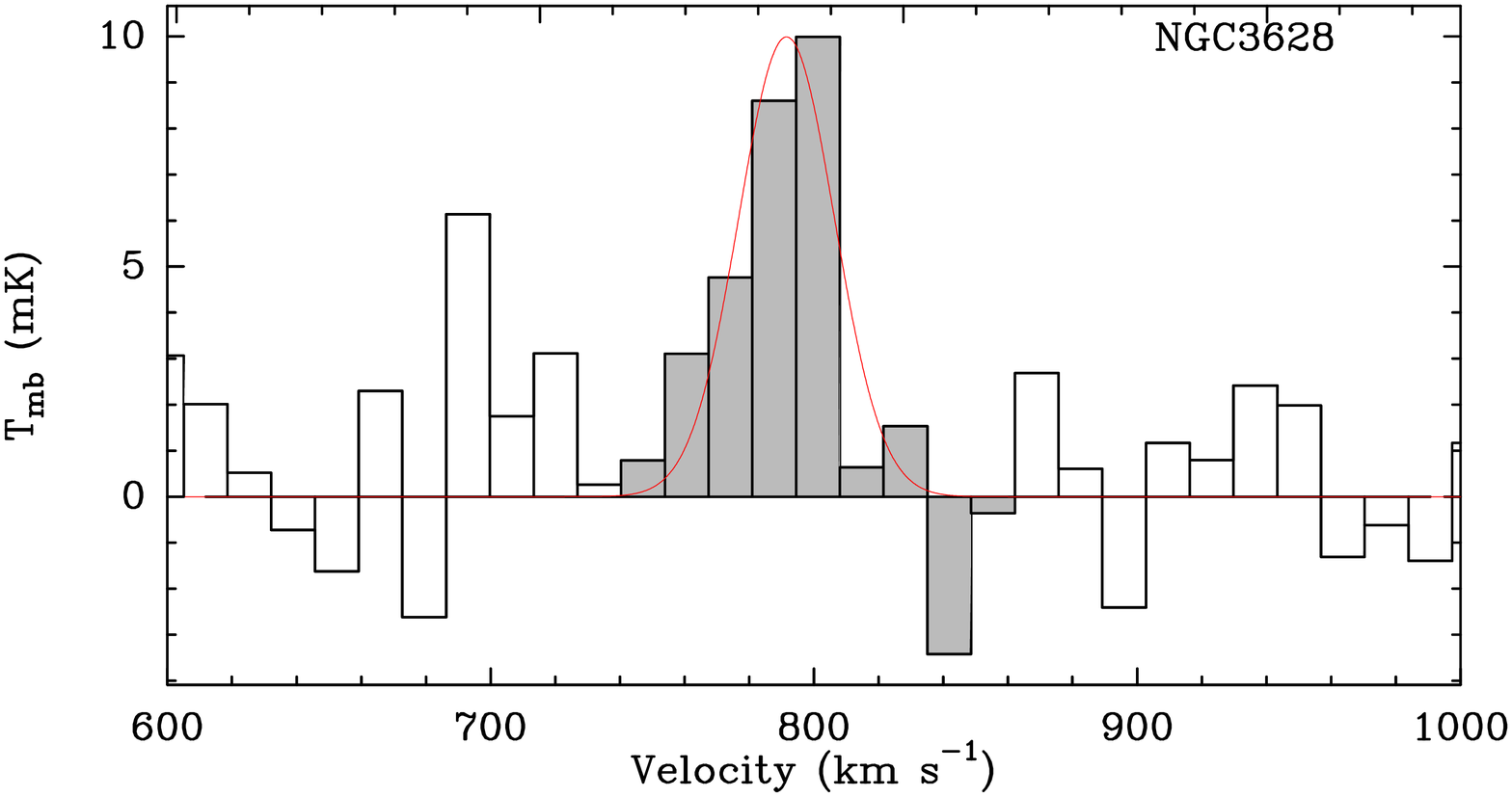}
\includegraphics[height=1.3in,width=2.2in]{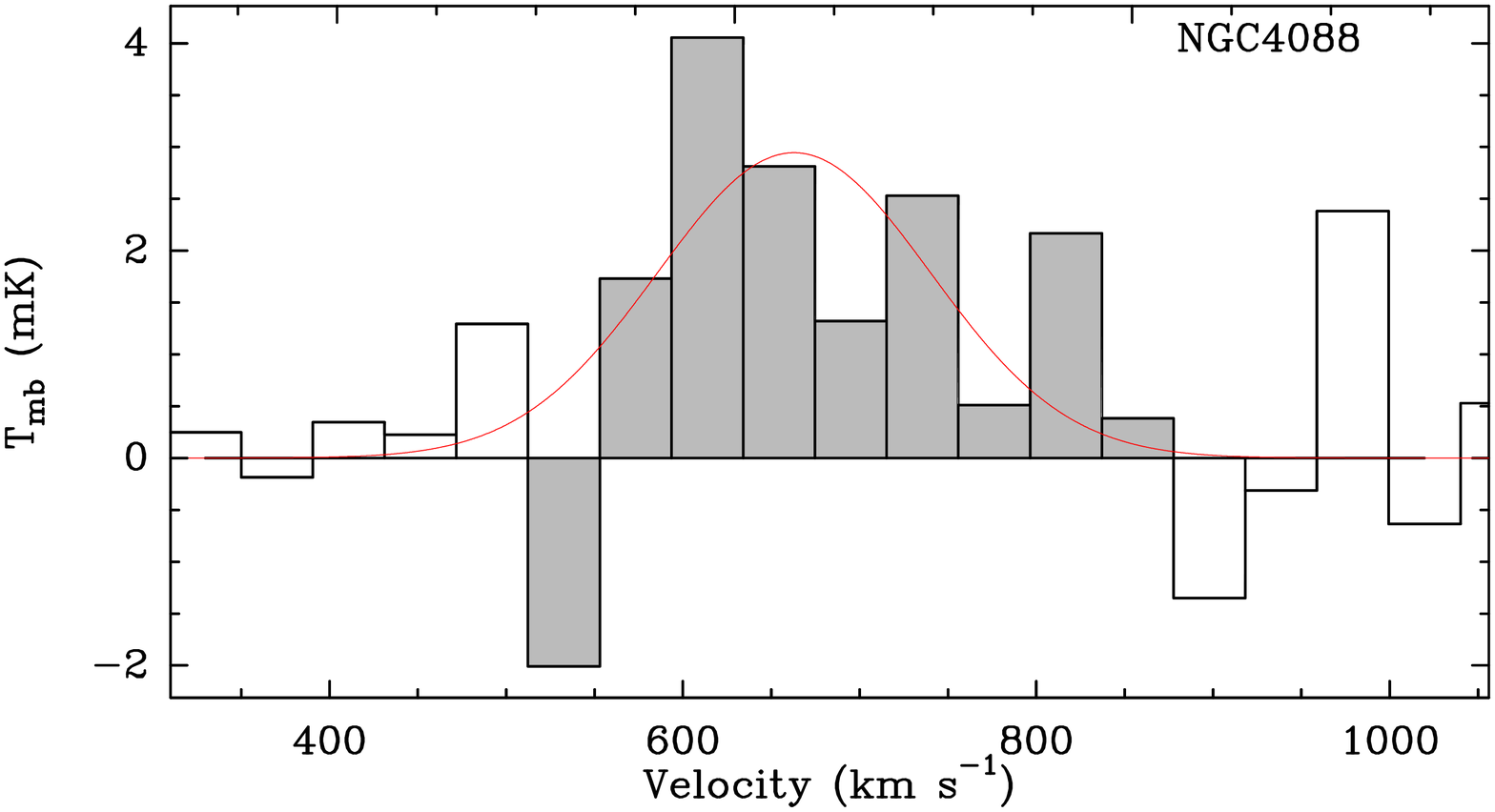}
\includegraphics[height=1.3in,width=2.2in]{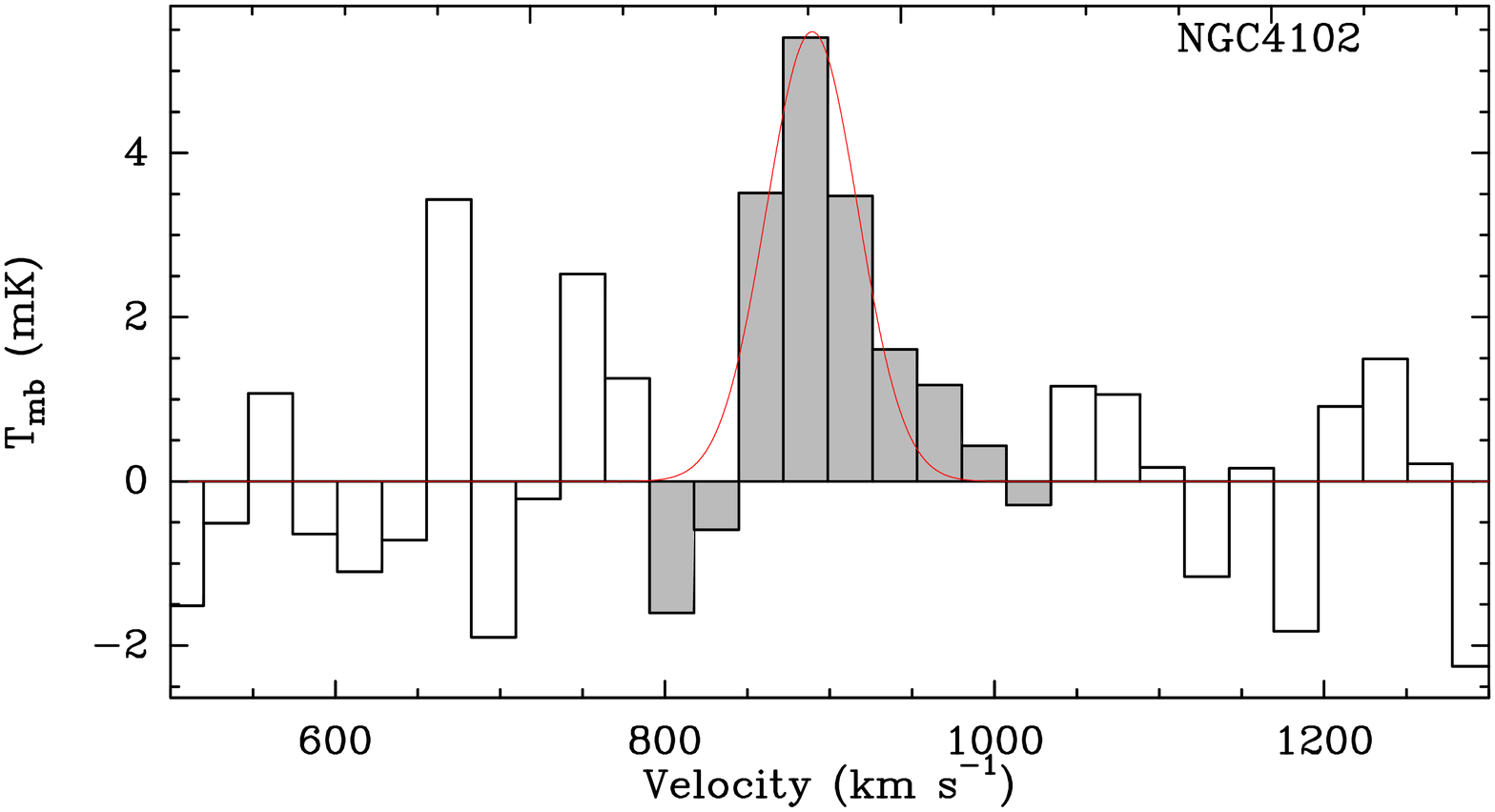}
\includegraphics[height=1.3in,width=2.2in]{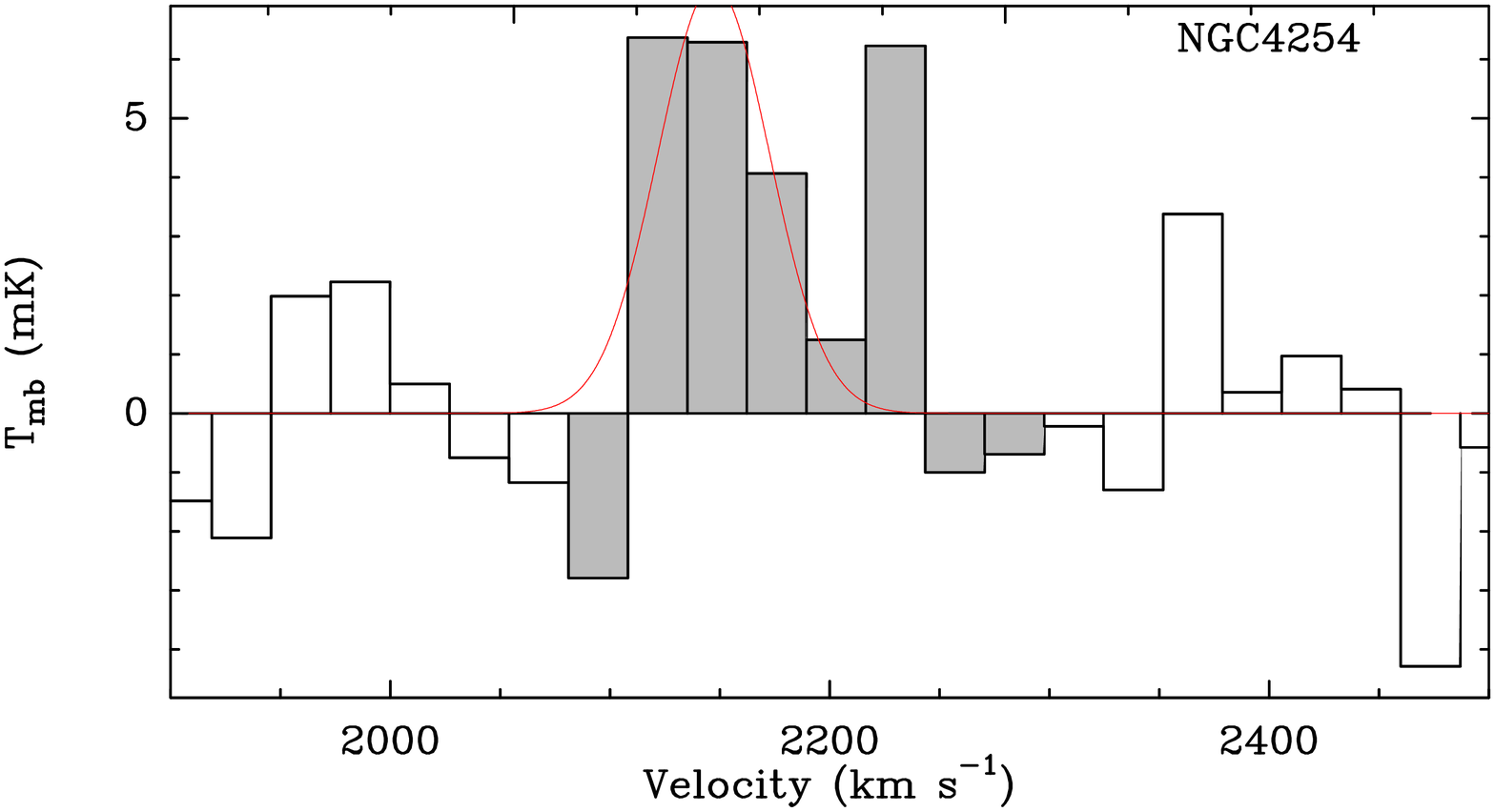}
\includegraphics[height=1.3in,width=2.2in]{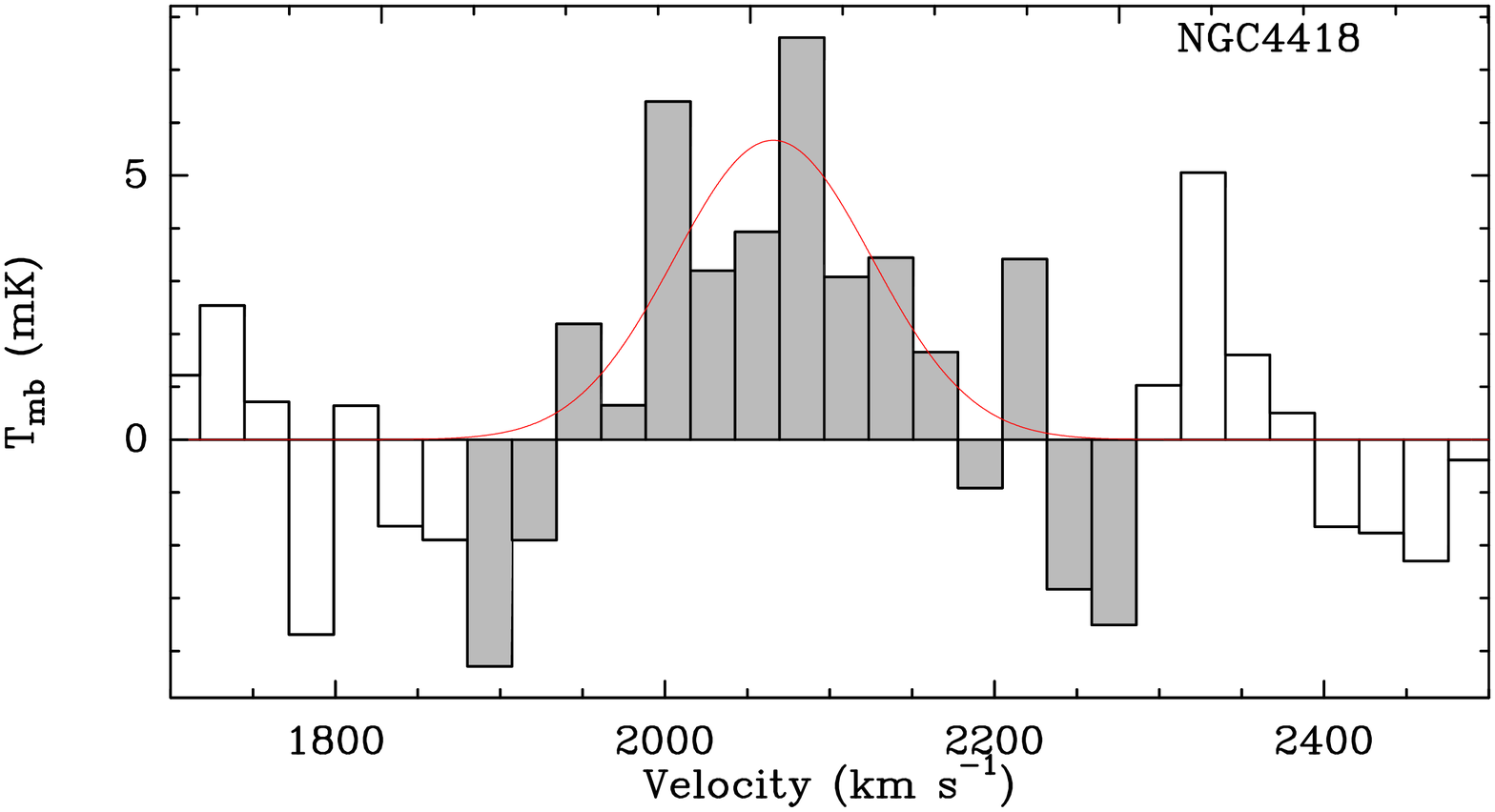}
\includegraphics[height=1.3in,width=2.2in]{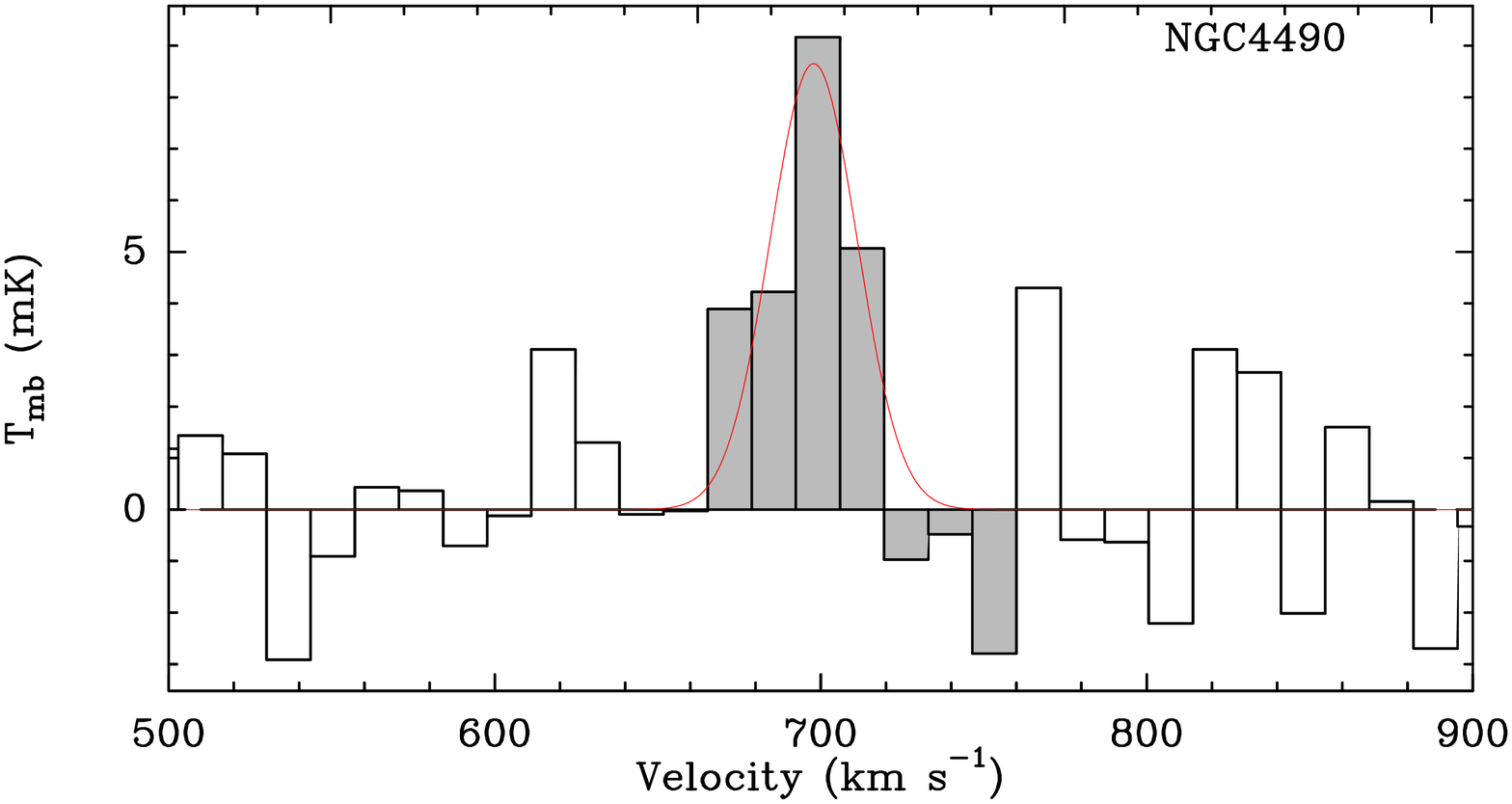}
\includegraphics[height=1.3in,width=2.2in]{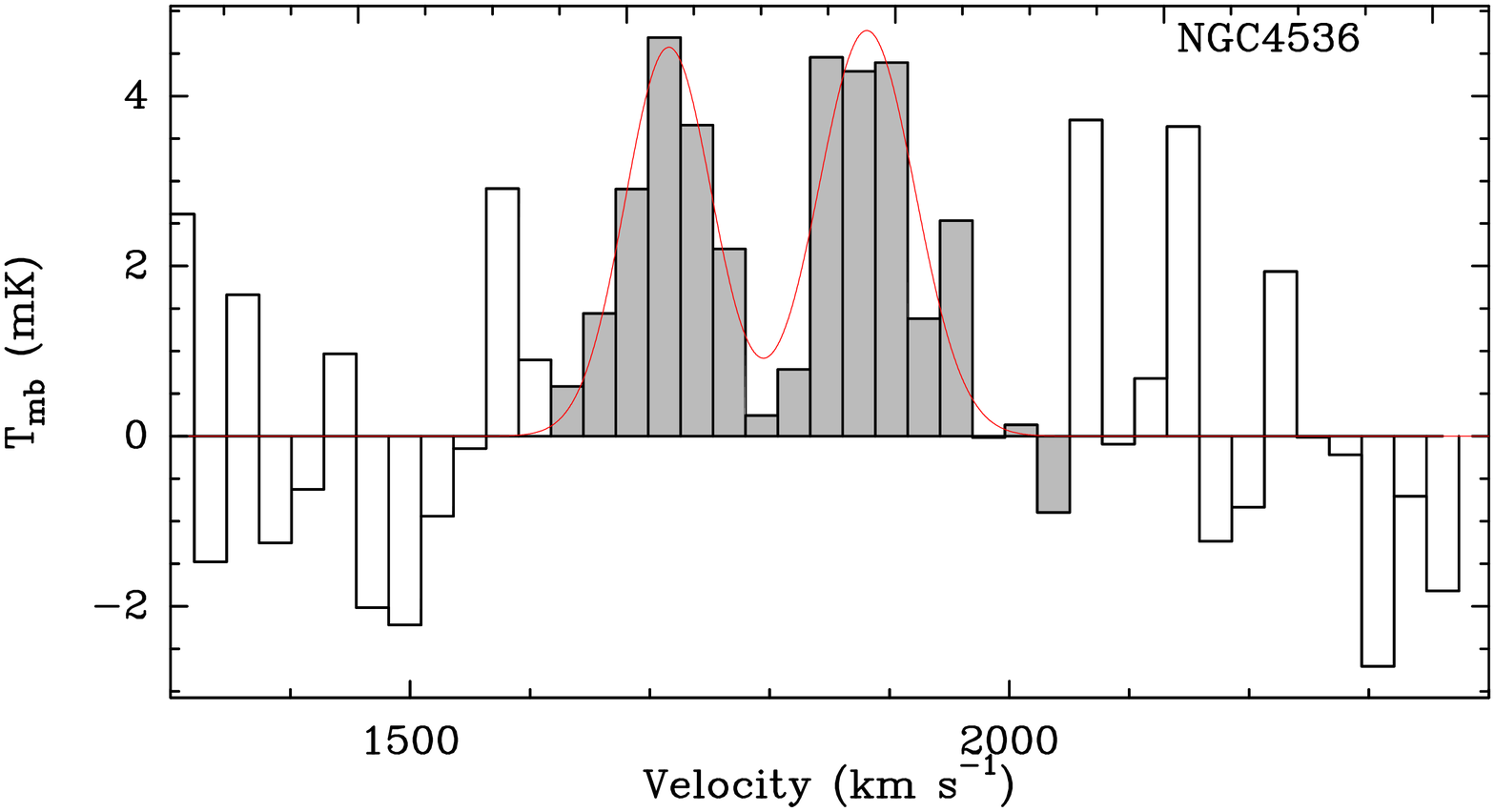}
\includegraphics[height=1.3in,width=2.2in]{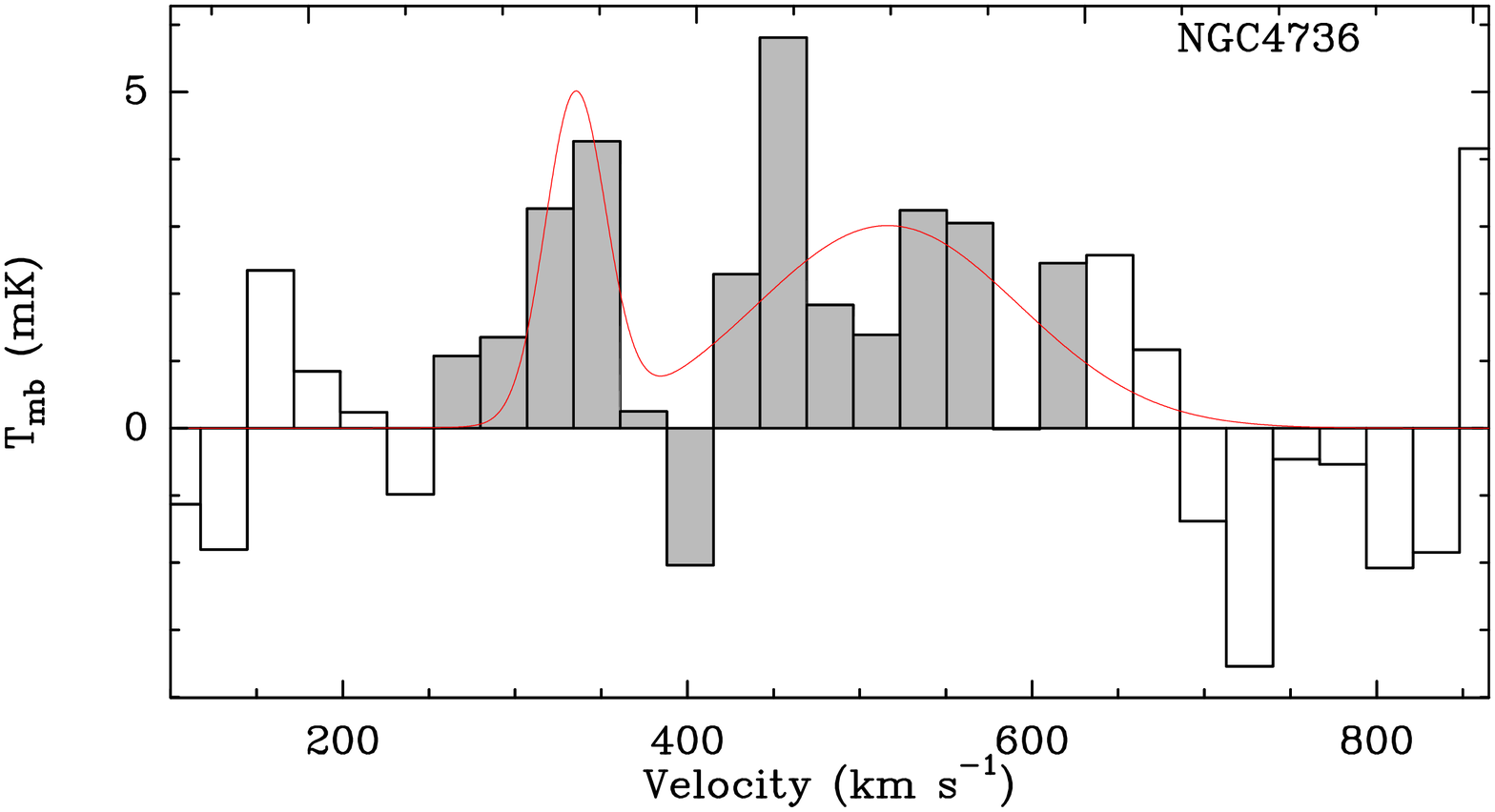}
\includegraphics[height=1.3in,width=2.2in]{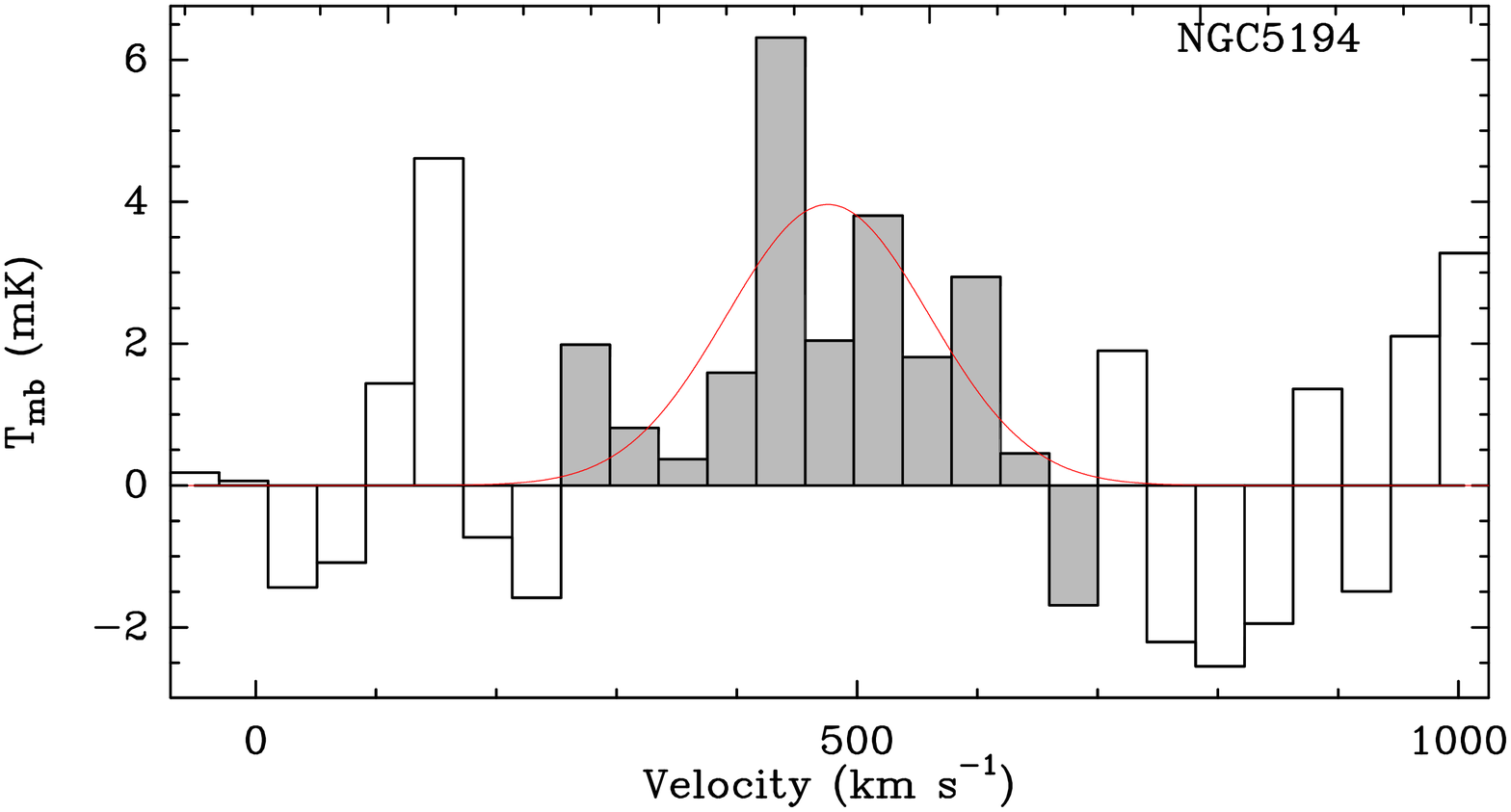}
\includegraphics[height=1.3in,width=2.2in]{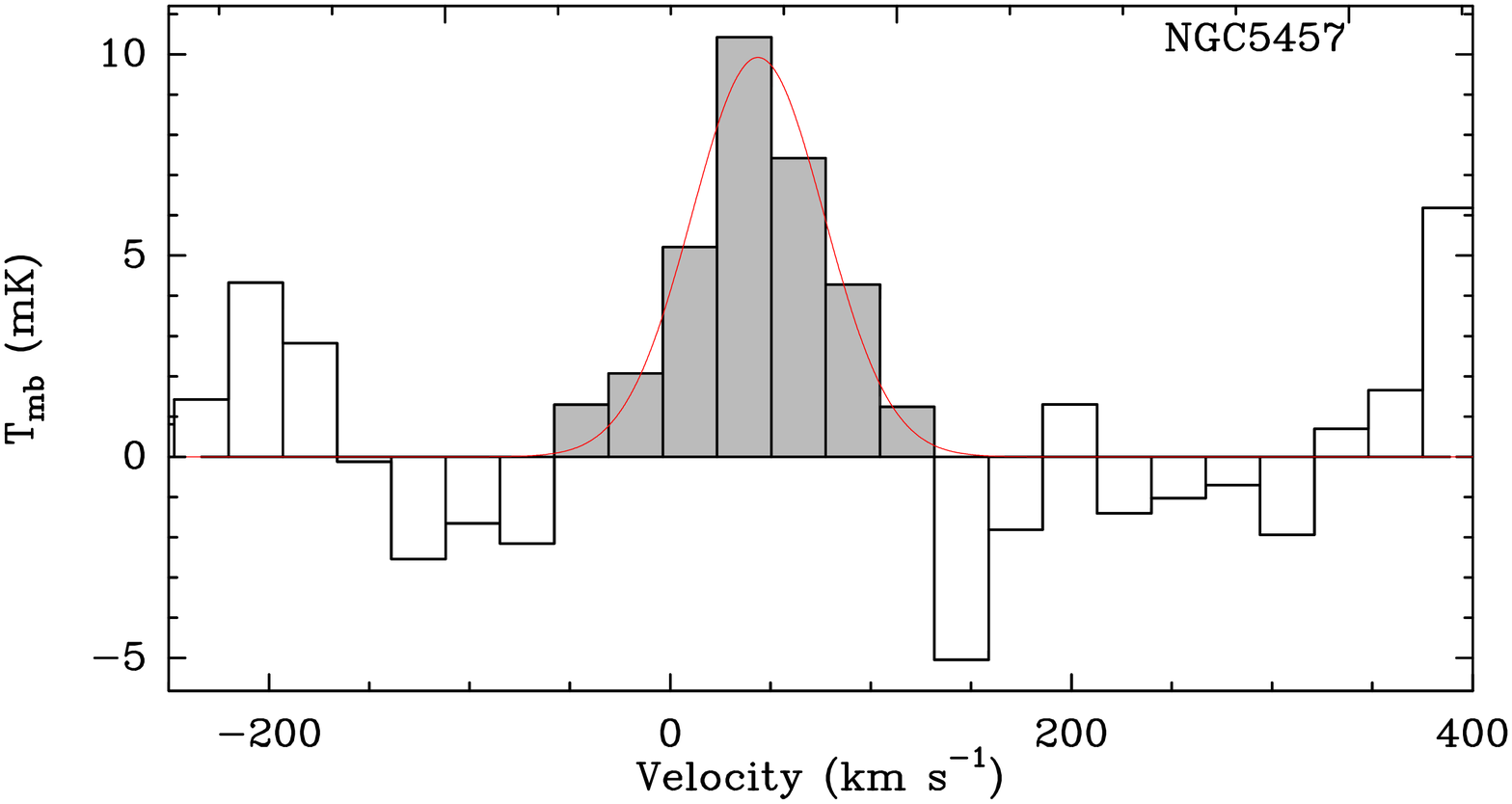}
\includegraphics[height=1.3in,width=2.2in]{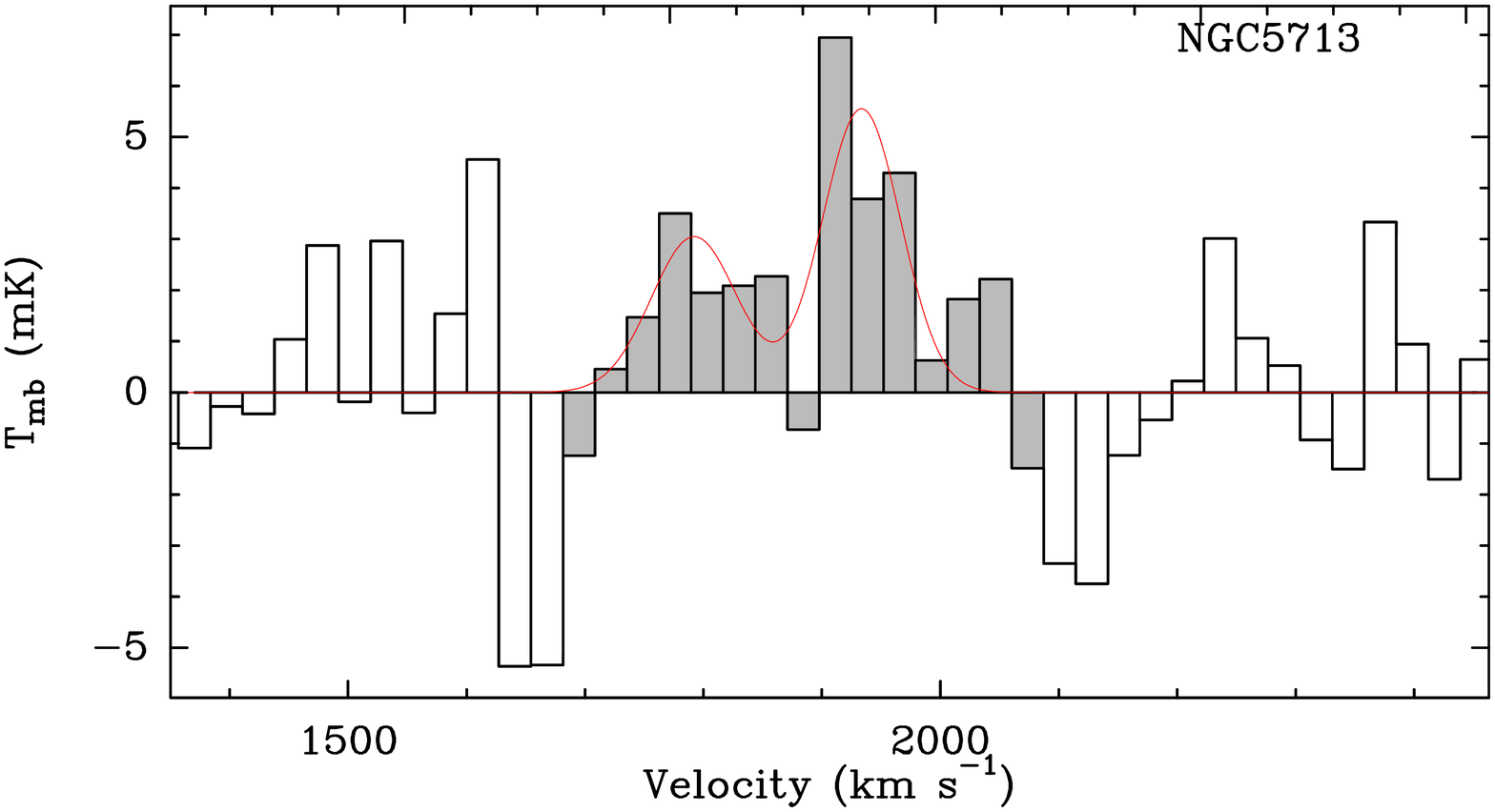}
\includegraphics[height=1.3in,width=2.2in]{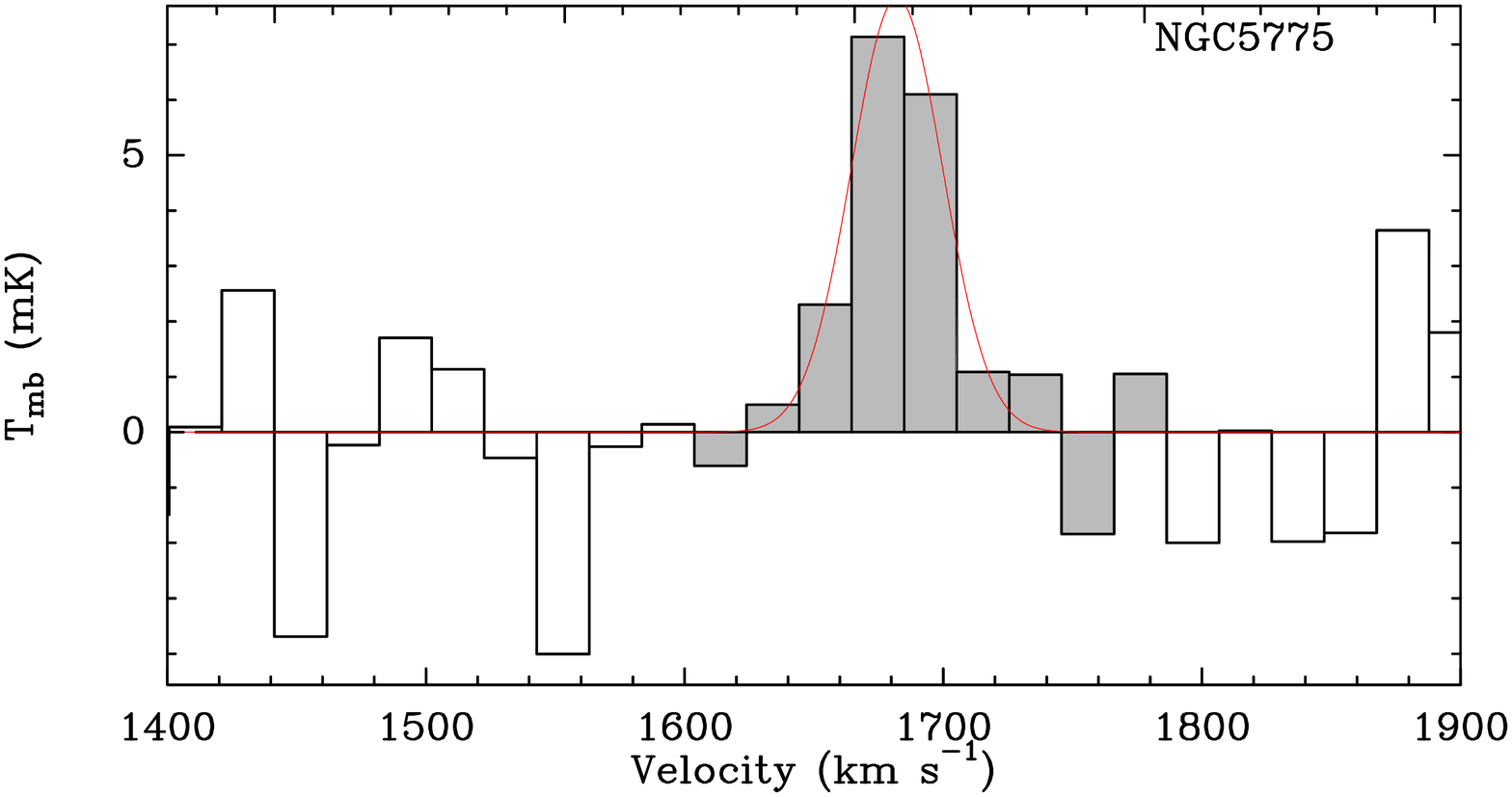}
\includegraphics[height=1.3in,width=2.2in]{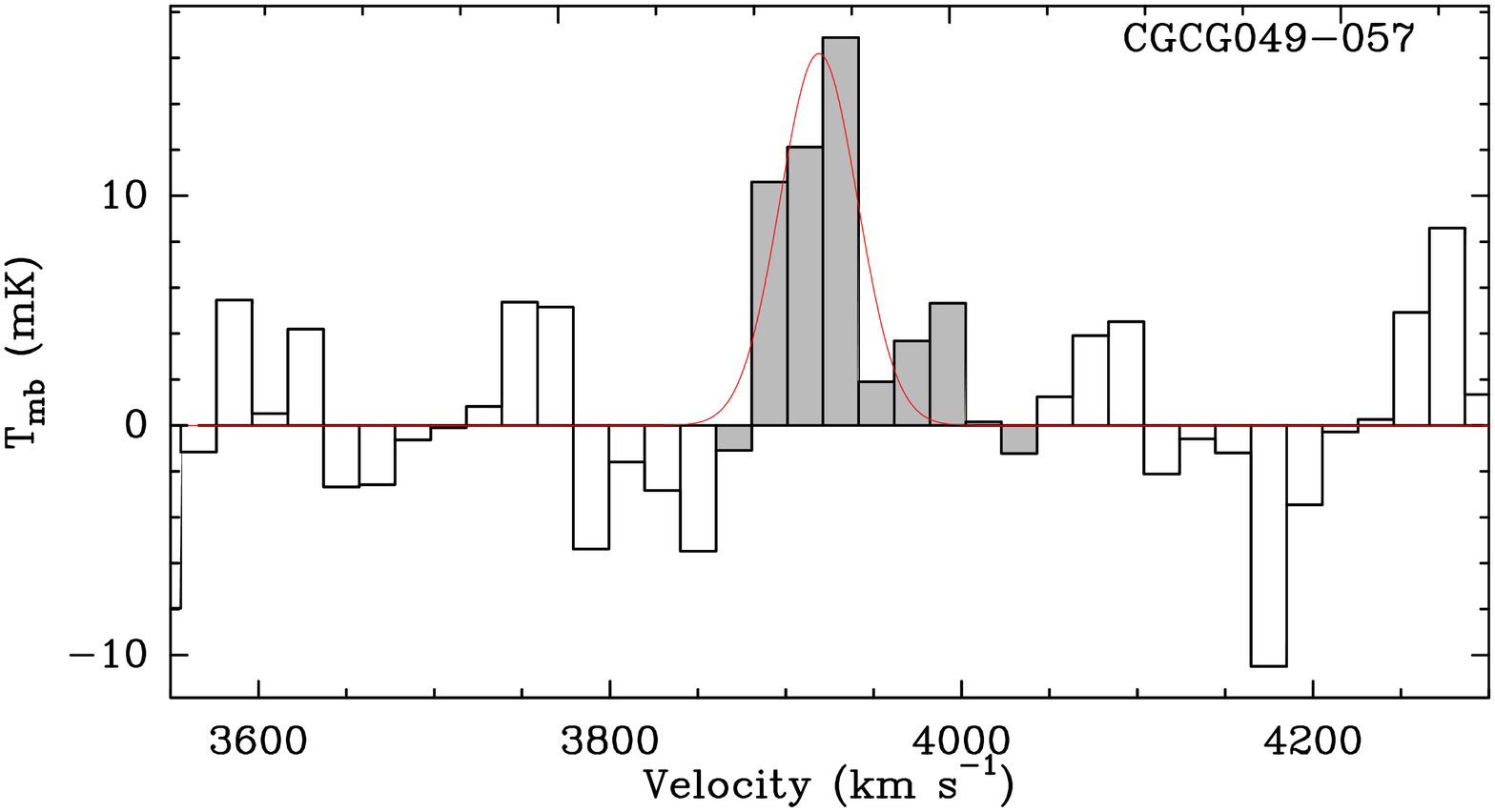}

  \end{figure*}
\label{fig:HCN 3-2_1}
\addtocounter{figure}{-1}

\begin{figure*} 
\addtocounter{figure}{1}      

 \subfigure{
  \includegraphics[height=1.3in,width=2.2in]{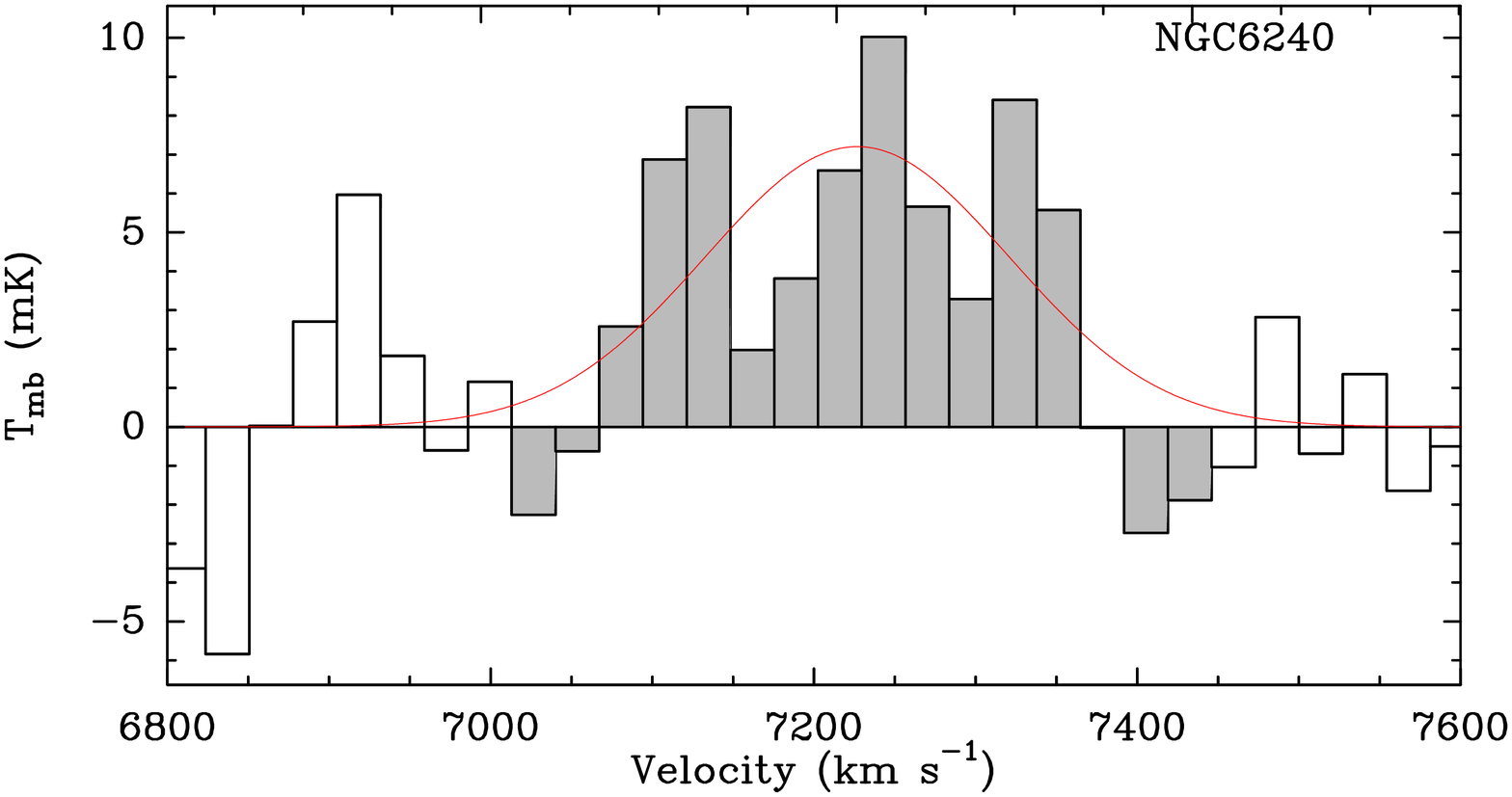}
   \includegraphics[height=1.3in,width=2.2in]{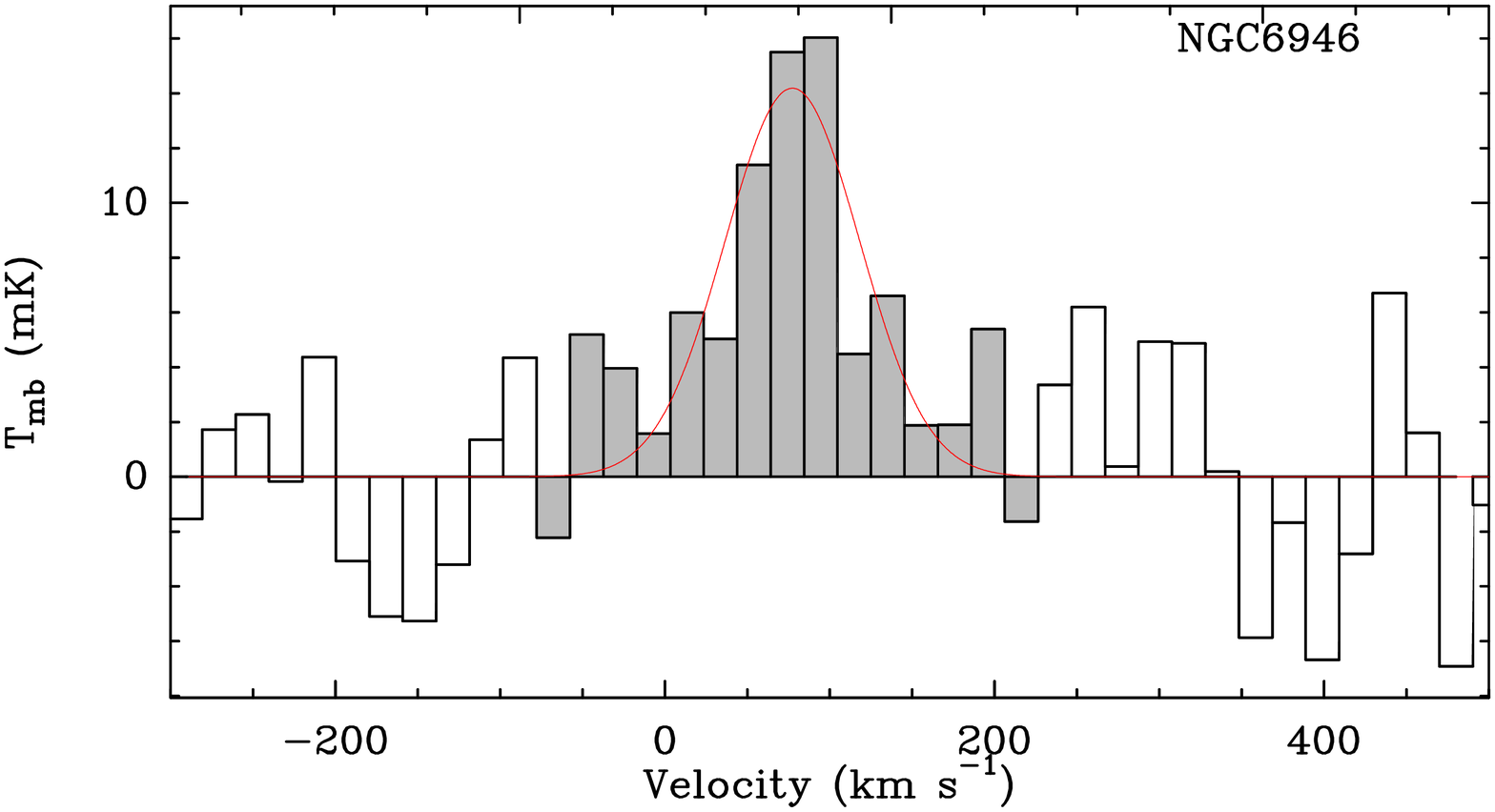}}
\caption{Spectra of detected HCN 3-2 in 23 galaxies (black histograms) and Gaussian fits (red lines). The velocity resolution is 27 km s$^{-1}$ in all cases}

\label{fig:HCN 3-2}
   
\end{figure*}


\subsection{Infrared Data}

To match the molecular emission, we measure the IR luminosity within the observing region of HCN 3-2, corresponding to the beam-size of 28$''$.
 We download the calibrated IR image data obtained using the Spitzer MIPS and Herschel PACS instruments from the NASA/IPAC Infrared Science Archive (IRSA). The data have been processed to level 2 for MIPS 24 $\mu$m and level 2.5 or 3 for PACS 70 $\mu$m, 100 $\mu$m, and 160 $\mu$m bands. For NGC~2903, NGC~4088, NGC~4490 and NGC~4414, we only  adopt Spitzer MIPS 24 $\mu$m image, because no Herschel data can be found. While the IR image data of all other sources are from Herschel PACS image. According to the method in \cite{Tan2018}, we calculate  the infrared flux densities from 24 $\mu$m to 160 $\mu$m.


\subsection{Infrared Luminosities and HCN 3-2 line Luminosity}

Using these IR data, we calculate the infrared luminosity of the region within the SMT beam size in each galaxy.
Based on the method in \cite{Galametz2013}, we estimate the total infrared luminosities using Spitzer MIPS and Herschel PACS luminosities. The  total IR luminosity within the beam-size is:

 \begin{equation}
\label{eq:pop}
 L_{\rm TIR}=\Sigma c_{i}vL_{v}(i)L_{\odot}  
\end{equation}

where $c_{i}$ is the calibration coefficients for various combinations of Spitzer and Herschel bands, $vL_{v}(i)$ is the resolved luminosity in a given band i in units of $L_{\odot}$. The errors include errors of photometry ($\sim$ 5\%),  the flux calibration error ($\sim$ 5\%) and the error of tracing TIR with a combined IR band ($\sim$ 25\%) \citep{Galametz2013}. We computed the HCN line luminosity using equation (2) in \cite{Gao2004b} for all galaxies. 


 \begin{equation}
\label{eq:pop}
L'_{\rm HCN} \approx \pi/(4ln2)\theta^2I_{HCN}d_L^2(1+z)^{-3}  K km s^{-1} pc^2  
\end{equation}


\section{Results}
\label{sec:result}

\subsection{The detection of HCN 3-2}

HCN 3-2 emission was detected in 23 galaxies from our sample of 37 local galaxies. Most of sources were detected at $>$ 5$\sigma$ level, except for NGC~5194, NGC~5457, NGC~3521, NGC~4088, NGC~5713, NGC~5775, and NGC~2146, which are at about 3$\sim$4$\sigma$ level. The distance, infrared luminosities of the objects and velocity integrate intensities of HCN 3-2 are presented in Table \ref{tabel 1}. The velocity  integrated intensities are consistent with the results of \cite{Bussmann2008} for seven sources overlapped with  their sample, including NGC~2146, NGC~2903, NGC~3034, NGC~3079, NGC~3628, NGC~4414, and NGC 6240. The HCN 3-2 spectrum of detected galaxies are presented in Figure \ref{fig:HCN 3-2}. The central velocity of HCN 3-2 is in agreement with that of the CO line in those detected galaxies \citep{Radford1991,Mauersberger1999,Meier2001,Israel2009,Mao2010,Costagliola2011,Usero2015}. Therefore, we use the CO line width to estimate the upper limits of velocity-integrated flux for the sources of non-detection in the HCN 3-2.

For both Mrk~231 and Mrk~273, HCN 3-2 is not detected, while the 3$\sigma$ upper limits are similar to the detection reported by \cite{Aalto2015} toward Mrk~ 231 and \cite{Aladro2018} toward Mrk~273. Therefore, the fluxes of HCN 3-2 in Mrk~231 \cite{Aalto2015}  and Mrk~273 \cite{Aladro2018}  are also used  for the correlation of  $L_{\rm IR}$ and $L'_{\rm HCN(3-2)}$. In addition,  two high-$z$ starbursts galaxies from \cite{Oteo2017} are also included,  which makes the sample with HCN 3-2 detection to be 41 galaxies. 

 
 \begin{figure*}[h]
\begin{center}
\includegraphics[width=5in]{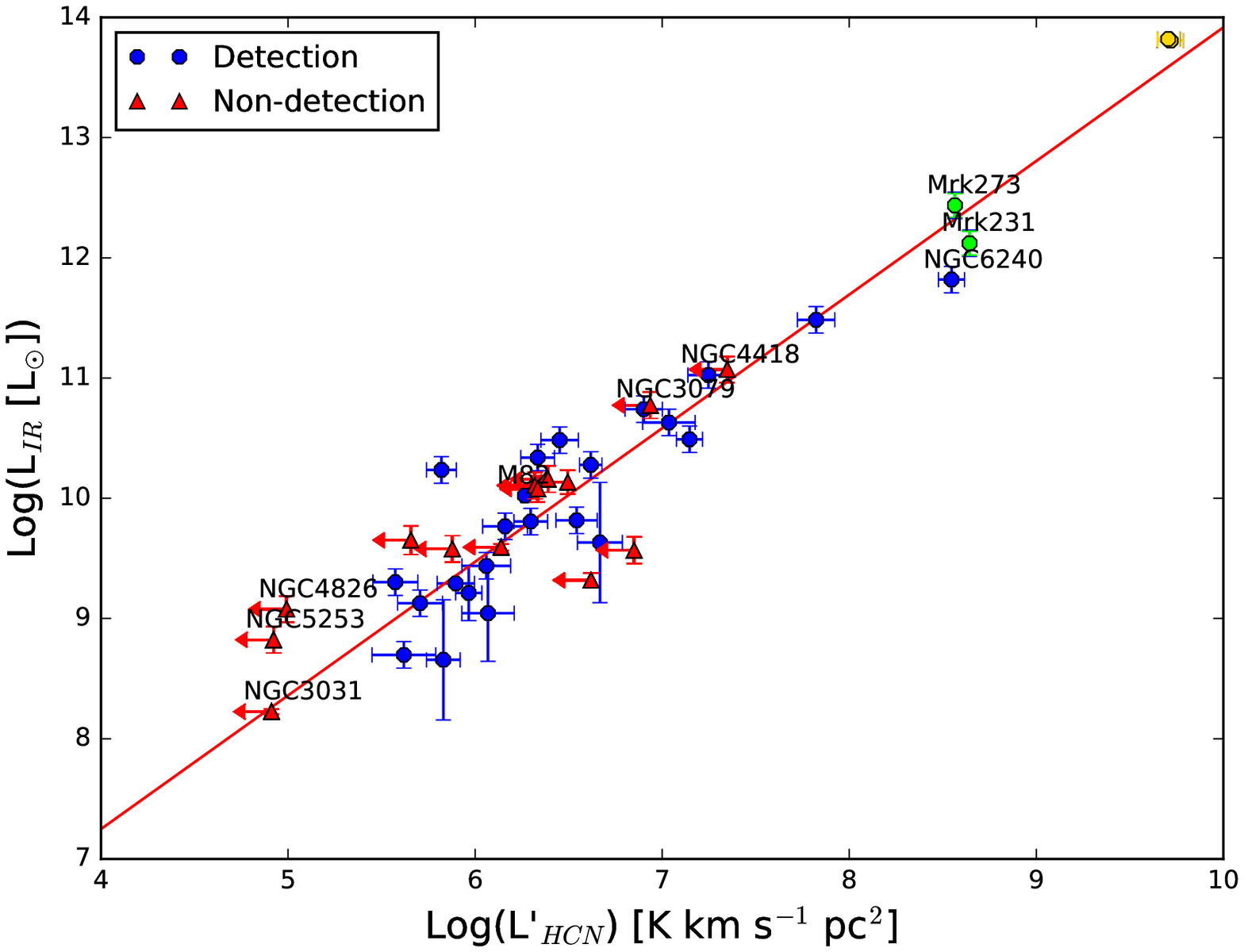}
\end{center}
\vspace*{-0.2 cm}
\caption{Correlations between the gas luminosity log($L'_{\rm HCN}$) and the IR luminosity log($L_{\rm IR}$). The blue circles are the sources with detected HCN 3-2 and the red triangle are the sources with non-detected HCN 3-2. The green circles are the Mrk~231 and Mrk 273, which is taken from \cite{Aalto2015} and \cite{Aladro2018}, respectively. The yellow circles are the two high-$z$ starbursts galaxies of SDP.9 and SDP.11, which is taken from \cite{Oteo2017}. The upper limits of HCN 3-2 for the non-detected galaxies are not adopted in the fitting.}
\label{fig:SFL_fit}
    \vskip5pt
\end{figure*}

 \subsection{The correlation of L$_{IR}$ and L$'_{HCN(3-2)}$}
 
Figure \ref{fig:SFL_fit} shows the relationship between the total infrared luminosity and HCN 3-2 luminosity using our new data and the data from the literatures. The infrared luminosity of the total galaxy has been corrected to the region within the SMT beam, as a proxy of SFR. In contrast with the result of a slope of 0.74 reported by \cite{Bussmann2008}, our results show that the L$_{IR}$ and L$'_{HCN}$ relation still follow the linear correlation. The sub-linear correlation between HCN 3-2 and IR in \cite{Bussmann2008} should be caused by that the infrared luminosities for nearby galaxies are not corrected to match the beam size of HCN 3-2 observation. The best-fit of result is: Log(L$_{IR}$)=1.11($\pm$0.06)Log(L$_{HCN}$)+2.80($\pm$0.42) and with a correlation coefficient of 0.91. The upper limits of HCN 3-2 for the non-detected galaxies are not adopted in the fitting. The $L_{IR}$ of sample galaxies span 5 orders of magnitude from about 10$^8L_\odot$ to 10$^{13}L_\odot$,
including various types of galaxies.


 \begin{figure*}[h]
\begin{center}
\includegraphics[width=3in]{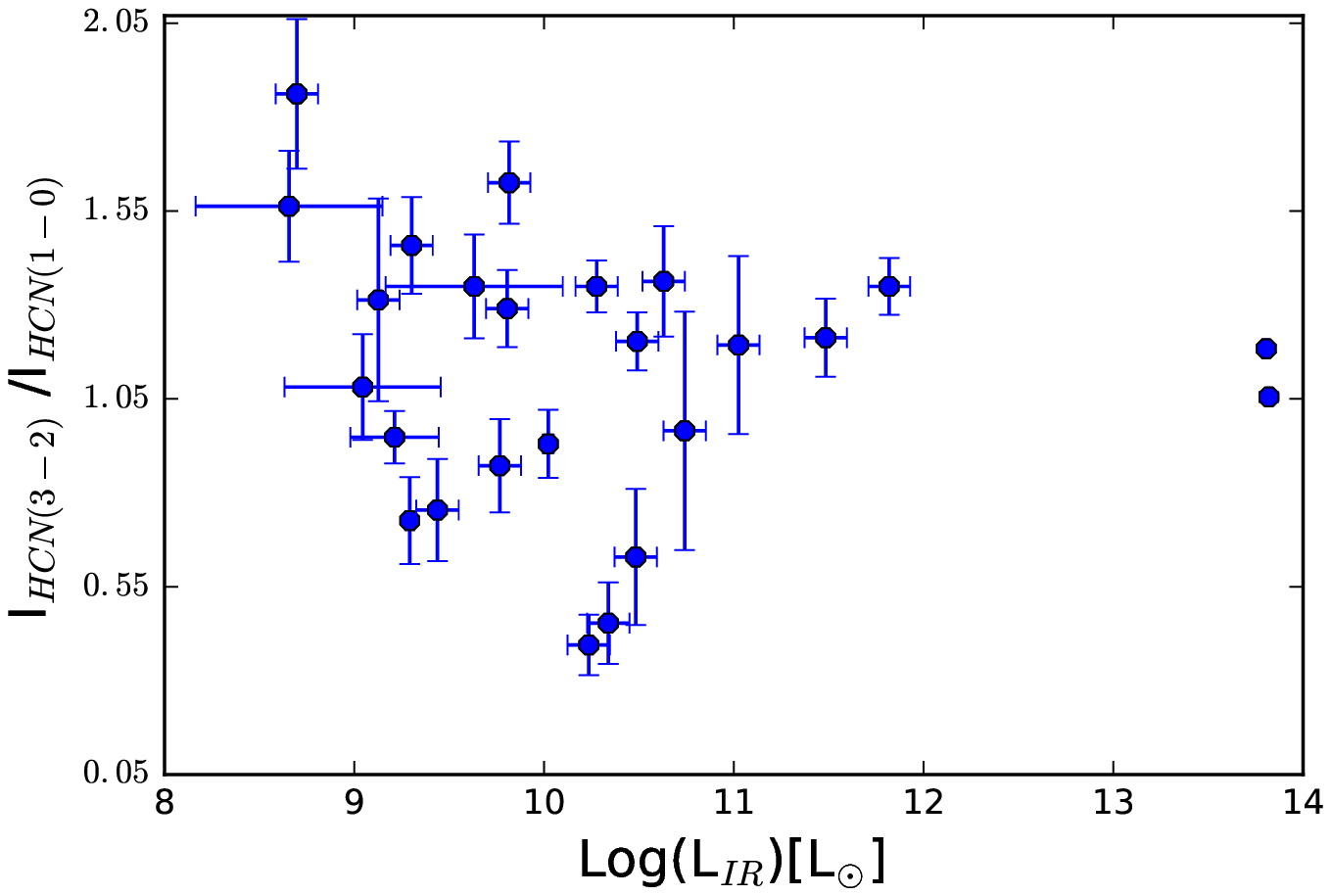}
\includegraphics[width=3in]{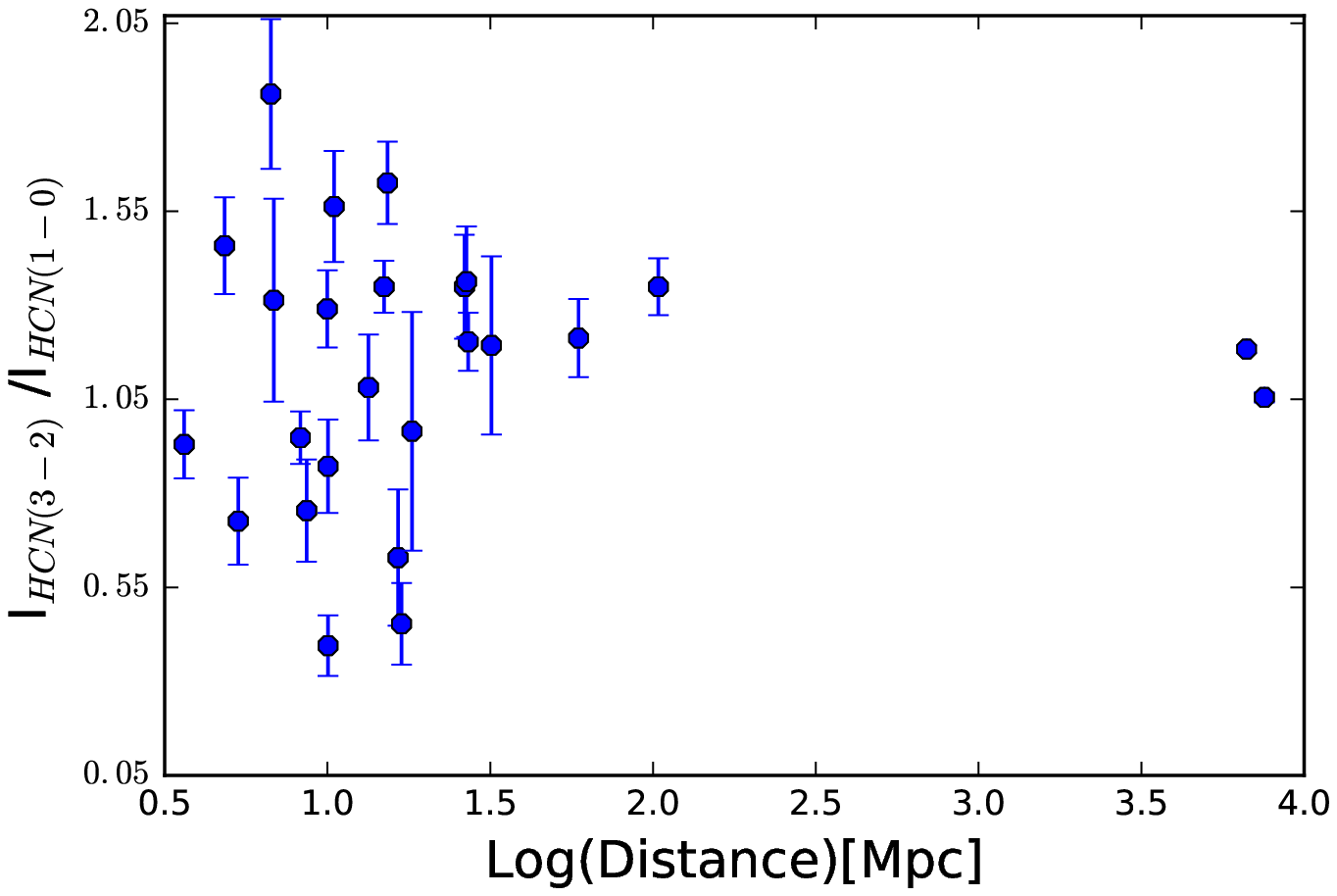}
\end{center}
\vspace*{-0.2 cm}
\caption{Left: integrated line ratios of HCN (3-2) and HCN (1-0) versus L$_{IR}$ . The HCN (1-0) data is observed with IRAM 30-m telescope in 2019, which is from the dense gas tracers survey toward the nearby galaxies by Wang et al. in 2019. Right: integrated line ratios of HCN (3-2) and HCN (1-0) versus luminosity distance for the same galaxies in the left panel.}
\label{fig:HCN3-2/LIR}
    \vskip5pt
\end{figure*}

\subsection{The line ratios versus L$_{IR}$}


The diagram of HCN 3-2/1-0 line ratio versus L$_{IR}$ is presented in Figure \ref{fig:HCN3-2/LIR},  which showed 
 large scatter in different galaxies. Such scatter indicates that excitation conditions of dense molecular gas vary in different galaxies. 
  There is no correlation between HCN 3-2/1-0 ratio and luminosity distance. This suggests that excitation conditions of dense molecular gas is independent of galaxy luminosity distance. Although both $L_{\rm IR}$-$L_{\rm HCN(3-2)}$ and $L_{\rm IR}$-$L_{\rm HCN(1-0)}$ show linear relation, different excitation conditions of dense gas would cause large uncertainty in estimating  dense gas mass for individual galaxy if only a single transition line of dense gas tracers is used.



\section{ Discussion}
\label{sec:discussion}

\subsection{The relation between luminosity of dense gas tracers and infrared luminosity}

The correlation of the dense gas content and the SFR in the local Universe was studied started from \cite{Solomon1992} and \cite{Gao2004b}, which find HCN emission and IR luminosity is closely related.  Although some observations of HCN 1-0\&3-2, CS 3-2 \citep{Baan2008,Bussmann2008,Gracia-Carpio2008} have shown a sub-linear relation in the local luminous and ultraluminous infrared galaxies (LIRGs and ULIRGs), which is consistent with the theoretical models \citep{Krumholz2007, Narayanan2008}. However these work did not properly account for the different sizes between the submilimeter dish aperture and the IR emission. 
Once the size is properly corrected, the linear correlation between luminosities of dense gas tracers and $L_{\rm IR}$ is confirmed \citep{Wang2011,Zhang2014,Greve2014,Liu2015,Kamenetzky2016,Shimajiri2017,Yang2017,Tan2018}. Even the high-$z$ galaxies also  follow such linear slope \citep{Oteo2017}.


The best fit of our data is  a slope of 1.11 with a correlation coefficient of 0.91. Note that, the infrared luminosities of entire galaxies have been corrected to the central 28$''$ regions in these galaxies, taking into account dense gas line fluxes are detected from the region of the beam size of the SMT 10-m telescope, according to the method using by \cite{Tan2018}, which adopts the prescription of \cite{Galametz2013}. Thus, there is no real exception of liner relation between  luminosity of dense gas tracers and infrared luminosity in galaxies.

The upper limits of HCN 3-2 for the non-detected galaxies are not adopted in the fitting.  Note that, NGC~5253 and NGC~4826 lie above the relation of $L_{\rm IR}$ and $L_{\rm HCN(3-2)}$. For dwarf starburst galaxy NGC~5253, with an extreme youth of starburst \citep{vandenBergh1980,Beck1996,Calzetti1997,Pellerin2007}, a high specific star formation rate \citep{Calzetti2015} and a high star formation efficiencies  were found in the central starburst of NGC~5253 \citep{Miura2018}. The extreme environments and some mechanism, like mechanical heating, non-star-forming ISM component might result in NGC~5253 to deviate from the relation.


As we know,  star formation in galaxies predominantly takes place 
in dense regions.  Lines of high dipole moment molecules, such as HCN, HCO$^+$, HNC and CS,  can be good tracers for such dense molecular gas. 
However, it  can be affected by various physical process. 
Such as, the abundance of HCN can be enhanced in XDR surrounding an AGN 
\citep{Costagliola2011,Privon2017}, while \cite{Izumi2016} reported that the mechanical heating
from a jet and shock could drive HCN enhancement around the vicinity of AGNs. 
 The uncertainties in the relation of L$_{IR}$ and L$'_{HCN}$ mainly result from the conversion 
 from dense molecular gas luminosity to the dense gas mass  and from the infrared luminosity to the 
 SFR \citep{Shimajiri2017}.  Due to the dense gas tracers are normally optically thick, there is a large uncertainty in estimating
the dense gas mass from a single transition line of a high dipole moment molecule, which is similar to the 
 issue of the CO-to-H$_2$ conversion factors \citep{Narayanan2012,Papadopoulos2007}.

\subsection{Excitation of HCN molecules}

With the new HCN 1-0 data towards a sample of 41 nearby galaxies using the IRAM-30-m telescope (Wang et al. in preparation),
we compare the ratio of HCN (3-2) and HCN (1-0) in different populating galaxies in the left panel of
Figure \ref{fig:HCN3-2/LIR}, since these were designed in this way
using different telescopes yet having similar telescope beam
sizes of these two observations. No systematic trend is found between HCN 3-2/HCN 1-0 
and L$_{IR}$. In terms of excitation properties of dense gas, the low ratios of HCN 3-2 /HCN 1-0 imply
that the HCN J=3-2 transition might be subthermal for M~82, NGC~3628 and NGC~4254. In addition,
 HCN 3-2/HCN 1-0 versus luminosity distance also show a large scatter (see Figure \ref{fig:HCN3-2/LIR}), which suggests that the excitation conditions of dense molecular gas is independent of luminosity distance.

Although most of observations have evidenced that both HCN 3-2 and 1-0 (n(H$_2$)$>$10$^{4}$ cm$^{-3}$) are linear related to the star formation of galaxies, the different excitation condition in different transitions could impact the estimation of dense gas mass for the individual galaxies. In the conversion from $L'_{\rm HCN}$ to the dense gas mass, we have better consider all transitions of HCN emission to reduce the uncertainties. 

\textbf{\subsection{Star formation efficiency of dense molecular gas}}

\begin{figure*}[h]
\begin{center}
\includegraphics[width=3in]{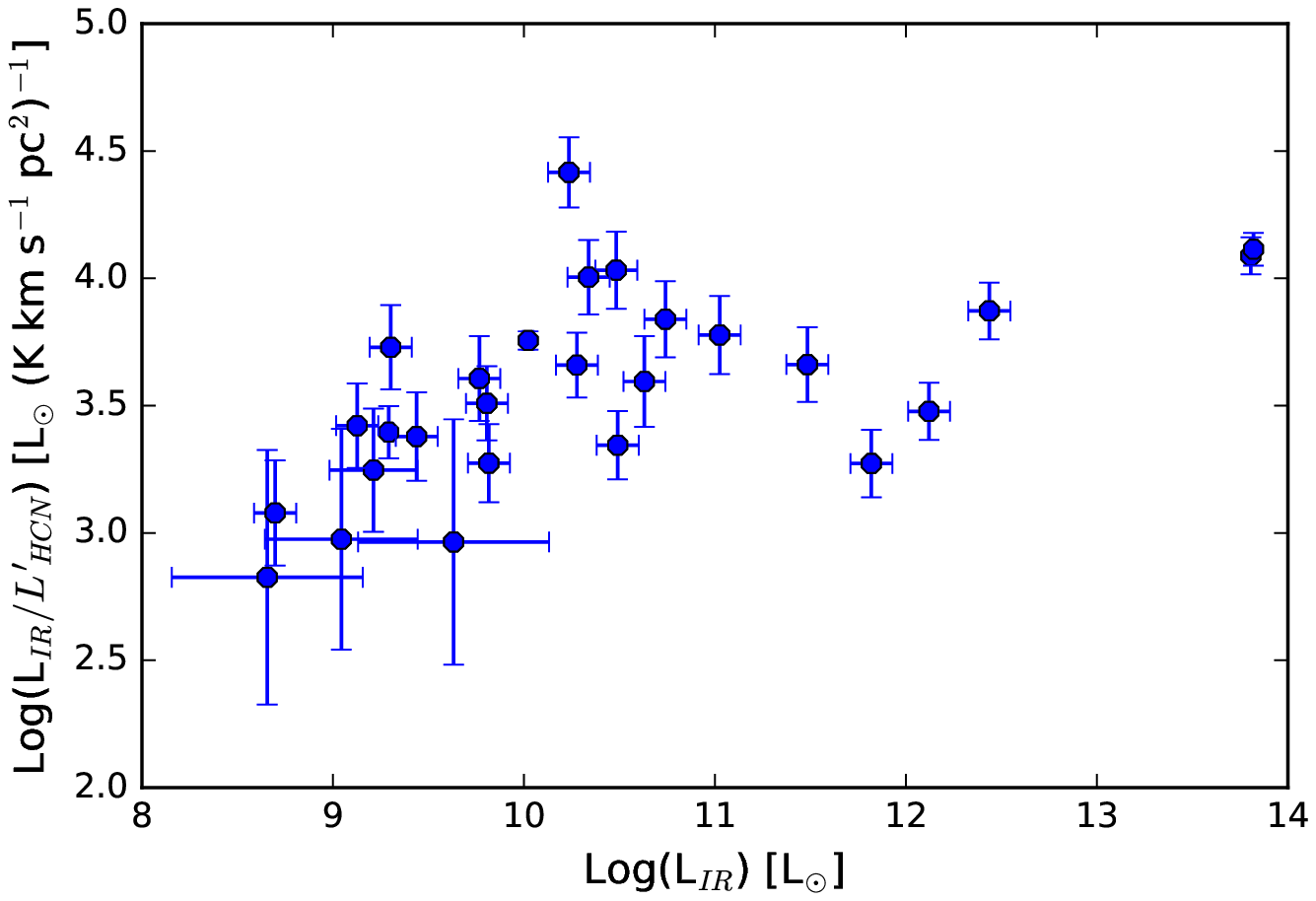}
\includegraphics[width=3in]{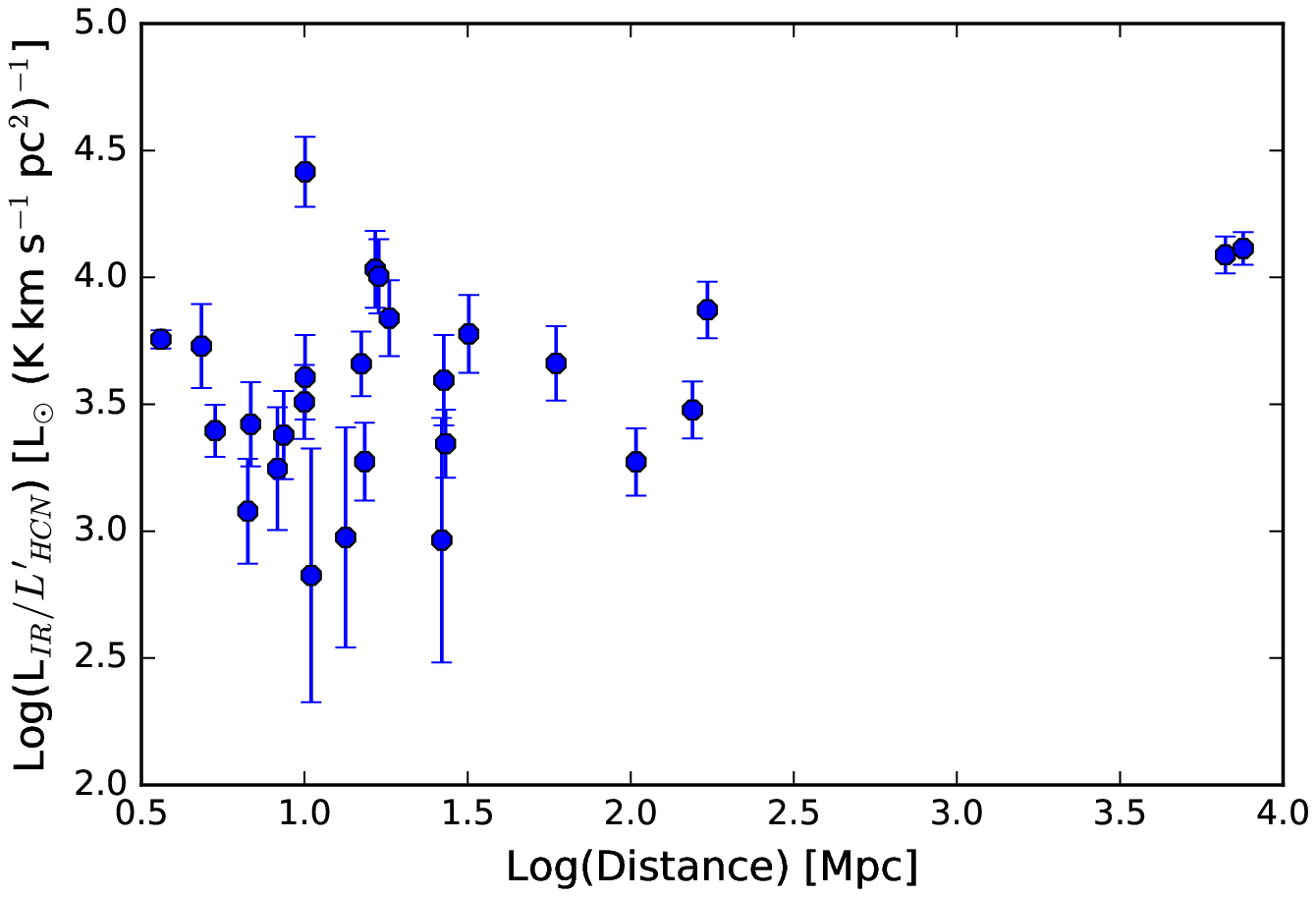}
\end{center}
\vspace*{-0.2 cm}
\caption{Left: the luminosity ratio of IR to HCN (3-2) as a function of $L_{\rm IR}$ for the detected galaxies and two high-$z$ galaxies from \cite{Oteo2017}. Right: the luminosity ratio of IR to HCN (3-2) as a function of luminosity distance for the same galaxies in the left panel.  }

\label{fig:LIR_LHCN}
    \vskip5pt
\end{figure*}

The relation between $L_{\rm IR}$/$L'_{\rm HCN}$ and $L_{\rm IR}$ can be interpreted as relation between the efficiency of star formation in the dense molecular gas (SFE) and SFR. We preform this relation toward the detected galaxies and two high-$z$ galaxies in the left panel of Figure \ref{fig:LIR_LHCN}. The ratio measured in different population of galaxies show little variations, which is consistent with the measure of the SFE of the dense gas traced by the HCN (1-0) in galaxies \citep{Gao2004a} and giant molecular clouds (GMCs) \citep{Wu2005,Chen2017} and HCN (4-3) in galaxies \citep{Tan2018}. The comparison of $L_{\rm IR}$/$L'_{\rm HCN}$ with luminosity distance reveals that SFE vary little (see the right panel of Figure \ref{fig:LIR_LHCN} ), which implies that SFE do not correlate with luminosity distance. The physical conditions of molecular gas in dense phase in a wide range of different galaxies might lead to the scatter in Figure \ref{fig:LIR_LHCN} \citep{Jackson1995A}. With the results from the literatures, various different transitions of HCN all indicate that SFE is almost constant independently of $L_{\rm IR}$.

\begin{figure*}[h]
\begin{center}
\includegraphics[width=5in]{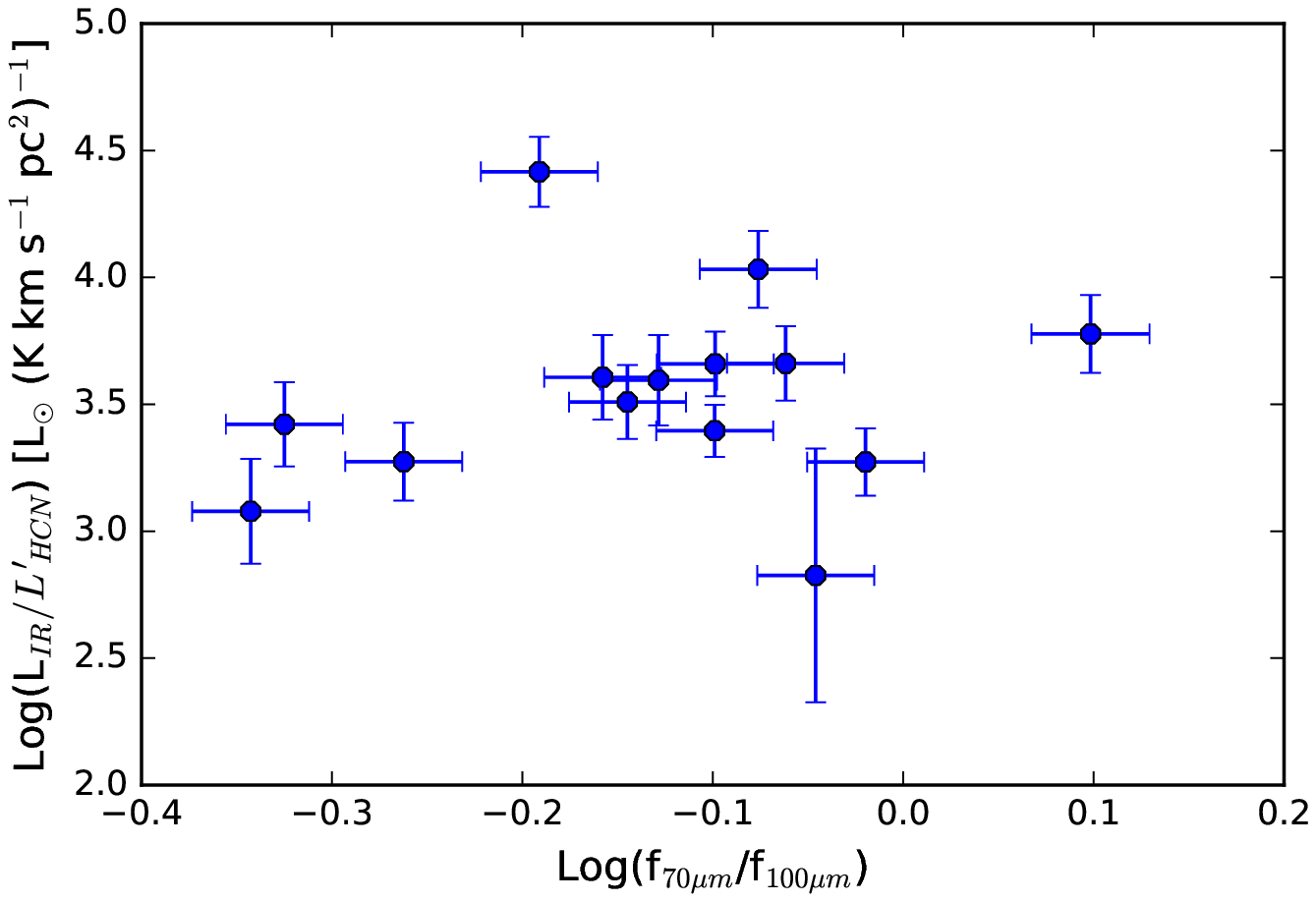}
\end{center}
\vspace*{-0.2 cm}
\caption{$L_{\rm IR}$/$L'_{\rm HCN (3-2)}$ as a function of $f_{70\mu m}/f_{100 \mu m}$ flux ratio for the galaxies
in our sample where we have both PACS 70$\mu m$ and 100$\mu m$ data. }

\label{fig:70um_100um}
    \vskip5pt
\end{figure*}

The warm-dust temperature can be traced by the $f_{ \rm 60 \mu m}/ f_{\rm 100 \mu m}$. Therefore, we compare the ratio of $L_{\rm IR}$/$L'_{\rm HCN}$ with $f_{\rm 70 \mu m}/f_{\rm 100 \mu m}$ for those galaxies with both PACS 70 $\mu$m  and 100 $ \mu$m data in Figure \ref{fig:70um_100um} instead of $f_{ \rm 60 \mu m}/ f_{\rm 100 \mu m}$.  
No significant correlation between the  L$_{IR}$/L$_{HCN}$ and the warm-dust temperature as traced by $f_{\rm 70 \mu m}/f_{\rm 100 \mu m}$ is found in our sample, although \cite{Tan2018} reported that there is a statistically significant correlation in the nearby star forming galaxies, and L$_{IR}$/L$_{HCN}$ is only weakly related with $f_{ \rm 60 \mu m}/ f_{\rm 100 \mu m}$ for the whole galaxies reported by \citep{Gao2004a}.

\section{Summary}
\label{sec:summary}

The observations of HCN 3-2 towards the nearby infrared galaxies with SMT 10-m
telescope are presented. Combined our data and the literature data of two high-z galaxies, we investigate the relation 
of the infrared luminosity ($L_{\rm IR}$) and the HCN luminosity ($L'_{\rm HCN}$) in different population galaxies. We
obtained the following results:

1. HCN 3-2 emission was detected in 23 out of 37 infrared bright galaxies. Most of sources were detected at $>$ 5$\sigma$ level, except for NGC 5194, NGC5457, NGC3521, NGC 4088, NGC 5713, NGC 5775, and NGC 2146, which is at about 3$\sim$4$\sigma$ level. The central velocity is consistent with the results of CO in those detected galaxies.

2.In contrast with the result of a slope of 0.74 reported by \cite{Bussmann2008}, the correlation of infrared emission (L$_{IR}$) and the luminosity of HCN (3-2) (L$'_{HCN}$) line emission measured in our sample is fitted with a slope of 1.11 and correlation coefficient of 0.91 after considering careful aperture corrections, which  follows the linear correlation for other dense gas tracers in literature established  for galaxies within the scatter.  


3. 
It is apparent that no systematic trend is found between the ratio of HCN(3-2)/HCN(1-0) 
and $L_{\rm IR}$. The the large scatter in the relation between the ratio of HCN(3-2)/HCN(1-0) versus $L_{\rm IR}$ indicates that dense gas masses estimated from the line luminosities of HCN $J$ = 1-0 and  $J$ = 3-2 should be treated with caution for individual galaxies.  In the conversion from L$'_{HCN}$ to the dense gas mass, we have better consider all transitions of HCN emission to reduce the uncertainties.

\section{Acknowledgements}

We are grateful to the staff of the SMT-10 m telescope and IRAM-30 m telescope for their kind help and support during out observation. This work is supported by the National Key R$\&$D Program of China (No. 2017YFA0402704), the Natural Science Foundation of China under grants of 11590783. This research has made use of the NASA/IPAC Extragalactic Database (NED), which is operated by the Jet Propulsion Laboratory, California Institute of Technology, under contract with the National Aeronautics and Space Administration. This work also benefited from the International Space Science Institute (ISSI/ISSI-BJ) in Bern and Beijing, thanks to the funding of the team ``Chemical abundances in the ISM: the litmus test of stellar IMF variations in galaxies across cosmic time'' (Principal Investigator D.R. and Z-Y.Z.). YG’s research is supported by National Key Research and Development Program of China (grant No. 2017YFA0402704), National Natural Science Foundation of China (grant Nos. 11861131007, 11420101002), Chinese Academy of Sciences Key Research Program of Frontier Sciences (grant No. QYZDJ-SSW-SLH008) and NSFC (grant No. U1731237).

\bibliographystyle{pasj}

\bibliography{SFL}

\end{document}